\documentclass{aa}

\usepackage{graphicx}
\usepackage{txfonts}
\usepackage{amsmath,amssymb}
\usepackage{mathrsfs}	% include it to use \mathscr font
\usepackage{mathtools}  % for subscripted mapsto
\usepackage{color}
\usepackage{breakurl}

\usepackage{natbib}
\bibpunct{(}{)}{;}{a}{}{,} % to follow the A&A style

\usepackage{longtable}
\usepackage{pdflscape} % for 'landscape' environment

\newcommand{\hi}{H{\sc i}}
\newcommand{\Msun}{M_\odot}
\newcommand{\Mstar}{M_\star}
\newcommand{\Mhalo}{M_{\rm halo}}

\newcommand{\Vc}{V_{\rm c}}

\newcommand{\Vobsi}{V_{{\rm obs},i}}
\newcommand{\Vgas}{V_{\rm gas}}

\newcommand{\Vdisc}{V_{\rm disc}}
\newcommand{\Vbulge}{V_{\rm bulge}}
\newcommand{\Udisc}{\Upsilon_{\rm disc}}
\newcommand{\Ubulge}{\Upsilon_{\rm bulge}}
\newcommand{\Vdm}{V_{\rm DM}}
\newcommand{\Vstar}{V_\star}
\newcommand{\fstar}{f_\star}
\newcommand{\fhi}{f_{\rm HI}}
\newcommand{\fhtwo}{f_{{\rm H}_2}}
\newcommand{\fb}{f_{\rm b}}
\newcommand{\fbaryons}{f_{\rm baryons}}
\newcommand{\Ustar}{\Upsilon_\star}
\newcommand{\Omegab}{\Omega_{\rm b}}

\newcommand{\Omegac}{\Omega_{\rm c}}

\defcitealias{SPARC}{LMS16}
\defcitealias{Katz+17}{K17}
\defcitealias{DuttonMaccio14}{DM14}

\begin{document}

\title{Peak star formation efficiency and no missing baryons in massive spirals}
\titlerunning{No missing baryons in massive spirals}
\authorrunning{L. Posti, F. Fraternali \& A. Marasco}

  \author{Lorenzo Posti\inst{1,2}\fnmsep\thanks{lorenzo.posti@astro.unistra.fr},
          Filippo Fraternali\inst{1}
          \and
          Antonino Marasco\inst{1,3}
          }
  \institute{Kapteyn Astronomical Institute, University of Groningen,
  			  P.O. Box 800, 9700 AV Groningen, the Netherlands
         \and
         Universit\'e de Strasbourg, CNRS UMR 7550, Observatoire astronomique de Strasbourg, 11 rue de l'Universit\'e, 67000 Strasbourg, France.
         \and
             ASTRON, Netherlands Institute for Radio Astronomy, Oude Hoogeveensedijk 4, 7991 PD, Dwingeloo, The Netherlands}
      \date{Received XXX; accepted YYY}

  \abstract{It is commonly believed that galaxies use, throughout the Hubble time, a very small fraction of the baryons associated to their dark matter halos to form stars. This so-called low "star formation efficiency" $\fstar\equiv \Mstar/\fb\Mhalo$, where $\fb\equiv\Omegab/\Omegac$ is the cosmological baryon fraction, is expected to reach its peak at nearly $L^\ast$ (at efficiency $\approx 20\%$) and decline steeply at lower and higher masses. We have tested this using a sample of nearby star-forming galaxies, from dwarfs ($\Mstar\simeq 10^7\Msun$) to high-mass spirals ($\Mstar\simeq 10^{11}\Msun$) with \hi~rotation curves and 3.6$\mu$m photometry. We fit the observed rotation curves with a Bayesian approach by varying three parameters, stellar mass-to-light ratio $\Ustar$, halo concentration $c$ and mass $\Mhalo$. We found two surprising results: 1) the star formation efficiency is a monotonically increasing function of $\Mstar$ with no sign of a decline at high masses, and 2) the most massive spirals ($\Mstar\simeq 1-3 \times 10^{11}\Msun$) have $\fstar\approx 0.3-1$, i.e.\ they have turned nearly all the baryons associated to their haloes into stars. These results imply that the most efficient galaxies at forming stars are massive spirals (not $L^\ast$ galaxies), they reach nearly 100\% efficiency and thus, once both their cold and hot gas is considered into the baryon budget, they have virtually no missing baryons. Moreover, there is no evidence of mass quenching of the star formation occurring in galaxies up to halo masses of $\Mhalo\approx {\rm a\, few}\times 10^{12}\Msun$.
             }
  \maketitle

\section{Introduction} \label{sec:intro}

In our Universe, only about one-sixth of the total matter is baryonic, while the
rest is widely thought to be in form of non-baryonic, collisionless, non-relativistic
dark matter \citep[e.g.][]{Planck18}. In the so-called standard $\Lambda$ Cold Dark
Matter ($\Lambda$CDM) paradigm, galaxies form within extended haloes of dark matter
that were able to grow enough to become gravitationally bound
\citep[e.g.][]{WhiteRees78}. In this scenario it is then reasonable to expect that
the amount baryons present in galaxies today is roughly a fraction
$\fb\equiv\Omega_{\rm b}/\Omegac= 0.188$ \citep[the ``cosmological baryon
fraction'', e.g.][]{Planck18} of the mass in dark matter. However, it was
realised that the total amount of baryons that we can directly observe in galaxies
(stars, gas, dust etc.) is instead at most only about 20$\%$ of the cosmological
value \citep[e.g.][]{PersicSalucci92, Fukugita+98}. This became known as the
``missing baryons'' problem and has prompted the search for large resevoirs of
baryons within the diffuse, multi-phase circumgalactic medium of
galaxies \citep{Bregman07,Tumlinson+17}.

Arguably the most important indicator of this issue is the so-called
\emph{stellar-to-halo mass relation}, which connects the stellar mass
$\Mstar$ of a galaxy to its dark matter halo of mass $\Mhalo$ \citep[see][for a recent review]
{WechslerTinker18}. This relation can be probed observationally through many
different techniques, e.g. galaxy abundance as a function of stellar mass
\citep[e.g.][]{ValeOstriker04,Behroozi+10,Moster+13}, galaxy clustering
\citep[e.g.][]{Kravtsov+04,Zheng+07}, group catalogues \citep[e.g.][]{Yang+08},
weak galaxy-galaxy lensing \citep[e.g.][]{Mandelbaum+06,Leauthaud+12}, satellite
kinematics \citep[e.g.][]{vandenBosch+04,More+11,WojtakMamon13} and internal galaxy
dynamics \citep[e.g.][hereafter \citetalias{Katz+17}]{Persic+96,McConnachie12,
Cappellari+13,DesmondWechsler15,Read+17,Katz+17}. Amongst all these determinations
there is wide consensus on the overall shape of the relation and, in particular, on the
fact that the ratio of stellar-to-halo mass $\fstar=\Mstar/\fb\Mhalo$, sometimes
called \emph{star-formation efficiency}, is a non-monotonic function of mass with
a peak ($\fstar\approx 0.2$) at $\Mhalo\approx 10^{12}\Msun$ (roughly the mass of the
Milky Way). This can be interpreted as galaxies of these characteristic mass
having been overall, during the course of their life, the most efficient at
turning gas into stars. And yet, efficiencies of the order of $20\%$ are still
relatively low, implying that most baryons are still undetected even
in these systems\footnote{Since molecular, atomic and ionized gas is typically
dynamically sub-dominant in $\Mstar>10^{10} \Msun$ galaxies.}.

Several works have suggested that the exact shape of the stellar-to-halo mass relation
depends on galaxy morphology \citep[e.g.][]{Mandelbaum+06,Conroy+07,
Dutton+10,More+11,Rodriguez-Puebla+15,Lange+18}, especially on the high-mass side
($\log\,\Mstar/\Msun\gtrsim 10$) where red, passive early-type systems appear to reside
in more massive halos with respect to blue, star-forming late-type galaxies.
This is intriguing, since it is suggesting
that galaxies with different morphologies likely followed different evolutionary
pathways that led the late-type ones, at a given M$_\star$, to live in relatively
lighter halos and to have a somewhat smaller fraction of missing baryons with respect
to early-type systems\footnote{
Blue galaxies also have typically larger reservoirs of cold gas with
respect to red ones. However, on average, the amount cold gas is sub-dominant
with respect to stars for $\Mstar>10^{10}\Msun$. \citep[e.g.][]{Papastergis+12}.
}. However, one of the main difficulties associated to these measurements is the paucity
of high-mass galaxies in the nearby Universe \citep[e.g.][]{Kelvin+14}, given that most
of the aforementioned observational probes use statistical estimates based on on large
galaxy samples.

In this paper we use another,
complementary approach to estimate the stellar-to-halo mass relation through
accurate modelling of the gas dynamics within spiral galaxies.
We use the observed \hi~rotation curves of a sample of regularly rotating,
nearby disc galaxies to fit mass models comprising of a baryonic plus a dark
matter component. We then extrapolate the dark matter profile to the virial
radius, with cosmologically motivated assumptions, to yield the halo mass.
A considerable advantage of this method is that each system can be studied
individually and halo masses, along with their associated uncertainties, can
be determined in great detail for each object.
We show that this approach leads to a coherent picture of the relation between
stellar and halo mass in late-type galaxies, which in turns profoundly affects
our perspective on the star-formation efficiency in the high-mass regime.

The paper is organised as follows: we present our sample and methodology to
derive stellar and halo masses in Section \ref{sec:method}; we describe our
results in Section \ref{sec:results} and we discuss them in detail in
Section \ref{sec:discussion}.

\section{Method} \label{sec:method}

Here we describe the data and methodology of our analysis. We adopt a standard $\Lambda$CDM
cosmology, with parameters estimated by the \cite{Planck18}. In particular, we use a Hubble
constant of $H_0=67.66$ km s$^{-1}$ Mpc$^{-1}$ and a cosmological baryon fraction of
$\fb\equiv\Omega_{\rm b}/\Omegac=0.188$.

\subsection{Data} \label{sec:data}

We use the sample of 175 disc galaxies with near-infrared photometry and \hi\ rotation curves
(SPARC) collected by \citet[][hereafter \citetalias{SPARC}]{SPARC}. This sample of spirals in
the nearby Universe spans more than 4 orders of magnitude in luminosity at 3.6$\mu$m and all
morphological types, from irregulars to lenticulars. The galaxies have been selected to have
extended, regular, high-quality \hi\ rotation curves and measured near-infrared photometry;
thus it is not volume limited. Nevertheless, it still provides a fair representation of the
population of (regularly rotating) spirals at $z=0$ and most importantly is best suited for our
dynamical study.

The \hi\ rotation curves are used as tracers of the circular velocity of the galaxies,
while the individual contributions of the atomic gas ($\Vgas$) and stars ($\Vstar$) to the
circular velocity are derived from the \hi\ and 3.6$\mu$m total intensity maps
respectively \citepalias[see][for further details]{SPARC}.
$\Vgas$ traces the distribution of atomic hydrogen, corrected for the presence of helium,
while the near-infrared surface brightness is decomposed into and exponential disc
($\Vdisc$) and a spherical bulge ($\Vbulge$). The contribution of the stars to the circular
velocity is then $V_\star^2=\Udisc\Vdisc^2 + \Ubulge\Vbulge^2$, given stellar
mass-to-light ratios of the disc ($\Udisc$) and bulge populations ($\Ubulge$)
respectively.

\subsection{Model} \label{sec:model}

We model the observed rotation curve as
\begin{equation} \label{eq:vcirc}
    \Vc = \sqrt{\Vdm^2 + \Vgas^2 + \Vstar^2}.
\end{equation}
Here $\Vdm$ is the dark matter contribution to the circular velocity and, for simplicity,
we have assumed that $\Ubulge=1.4\Udisc$, as suggested by stellar population synthesis models
\citep[e.g.][]{SchombertMcGaugh14}, thus $\Vstar^2 = \Udisc \left(\Vdisc^2 +
1.4 \Vbulge^2\right)$. In Appendix \ref{appendix} we explore the effect of fixing different
mass-to-light ratios $\Udisc$ and $\Ubulge$ for disc and bulge respectively: our findings on
the stellar-to-halo mass relation do not change significantly if we assume $\Udisc=0.5$ and
$\Ubulge=0.7$, for which the scatter of the baryonic Tully-Fisher relation is minimised
\citep{Lelli+16a}.

The dark matter distribution is modelled as a \citet[][hereafter NFW]{NFW}
spherical halo, which is characterised by a dimensionless concentration parameter ($c$)
and the halo mass ($\Mhalo$), which we take as that within a radius enclosing 200 times
the critical density of the Universe. Thus our rotation curve model has three free parameters:
$\Mhalo$, $c$ and $\Ustar$.

We compute the posterior distributions of these parameters with a Bayesian approach.
We define a standard $\chi^2$ likelihood $\mathcal{P}$, given the data
$\theta$, as
\begin{equation} \label{eq:likelihood}
\begin{split}
        \chi^2 &=  -\ln{\mathcal{P}(\theta|\Mhalo,c,\Udisc)} \\
        &= \sum_{i=0}^{N} \frac{1}{2}\left[\frac{\Vobsi -
        \Vc(R_i|\Mhalo,c,\Udisc)}{\sigma_{\Vobsi}}\right]^2
\end{split}
\end{equation}
where $\Vobsi$ is the $i$-th point of the observed rotation curve at radius $R_i$ and
$\sigma_{\Vobsi}$ is its observed uncertainty. The posterior distribution of the three
parameters is then given by Bayes' theorem
\begin{equation} \label{eq:bayes}
    \mathcal{P}(\Mhalo,c,\Udisc|\theta) \propto
        \mathcal{P}(\theta|\Mhalo,c,\Udisc)\,\mathcal{P}(\Mhalo,c,\Udisc)
\end{equation}
where $\mathcal{P}(\Mhalo,c,\Udisc)$ is the prior. We sample the posterior
with an affine-invariant Markov Chain Monte Carlo method \citep[MCMC, in particular, we
use the \texttt{python} implementation by][]{emcee}.

We use a flat prior on the stellar mass-to-light ratio $\Udisc$ limited to a reasonable
range, $0.01 \lesssim \Udisc \lesssim 1.2$, which encompasses estimates obtained with
stellar population models \citep[][]{Meidt+14, McGaughSchombert14}.
In a $\Lambda$CDM Universe the halo mass and concentration are well known to be
anti-correlated. Thus, in order to test whether standard $\Lambda$CDM haloes can be
used to fit galaxy rotation curves and then yield a stellar-to-halo mass relation, for the
halo concentration we assume a prior which follows the $c-\Mhalo$ relation as estimated in N-body
cosmological simulations \citep[e.g.][hereafter \citetalias{DuttonMaccio14}]{DuttonMaccio14}:
for each $\Mhalo$, the prior on $c$ is lognormal with mean and uncertainty given by the $c=c(\Mhalo)$
of \citetalias{DuttonMaccio14} (their Eq.~8). The prior on the dark matter halo mass $\Mhalo$
is, instead, flat over a wide range: $6 \leq \log\,\Mhalo/\Msun \leq 15$.

A non-uniform prior on the halo concentration is needed to infer reasonable constraints
on the halo parameters (see e.g. \citetalias{Katz+17}). The reason for this is that the
\hi~rotation curves do not typically extend out enough to probe the region where the NFW
density profile steepens, thus yielding only a weak inference on $c$. The $\Lambda$CDM-motivated
prior on the $c-\Mhalo$ relation proves to be enough to well constrain all the model
parameters. Furthermore, we notice that the \citetalias{DuttonMaccio14} $c-\Mhalo$
relation does not distinguish between haloes hosting late-type or early-type galaxies,
so we use it under the assumption that it provides a reasonable description of the
correlation for the haloes where late-type galaxies form.
We summarise in Table~\ref{tab:priors} our choice of priors.

%------ TAB. 1 --------------------------------------------------------------------------------
\begin{table}
\caption{Priors of our model. $\mathcal{P}(\Mhalo,c,\Ustar)$ in Eq.~\ref{eq:bayes} is given
by the product of the three terms.}
\label{tab:priors}
\begin{center}
\begin{tabular}{lcc}
\hline\hline \\[-.2cm]
Parameter & Type &  \vspace{.1cm}\\
\hline\hline \\[-.2cm]
$\Ustar$ &  uniform & $0.01 \leq \Ustar \leq 1.2$ \\
$\Mhalo$ &  uniform & $6 \leq \log\Mhalo/\Msun \leq 15$ \\
$c$ &  lognormal & $c-\Mhalo$ from \citetalias{DuttonMaccio14} \\
\hline
\end{tabular}
\end{center}
\end{table}

\section{Results} \label{sec:results}

We modelled the rotation curves and we have measured the posterior distributions of
$\Udisc, \Mhalo$ and  $c$ for all the 158 SPARC galaxies with inclination on the sky
larger than 30 degrees -- since for nearly face-on systems the rotation curves are very
uncertain. For each parameter, we define the ``best-value'' to be the median of the
posterior distribution and its uncertainty as the 16th - 84th percentiles.
In Appendix~\ref{appendix} we provide in tabular form all the measurements and
uncertainties, together with the value of the likelihood associated to the best model
(Table \ref{tab:results}). We also present the full rotation curve decomposition for
one case as an example (NGC 3992, Figure~\ref{fig:curve_decomp}), while we make
available the plots of all the other galaxies online at
\url{http://astro.u-strasbg.fr/~posti/PFM19_fiducial_fits/}.

Unsurprisingly, we find that our model typically does not give very stringent constraints
on the stellar mass-to-light ratio, with only 84 (68) galaxies having an uncertainty on
$\Ustar$ smaller than  50$\%$ (30$\%$). In these cases, which are mostly for
$\Mstar>10^{10}\Msun$ where the signal-to-noise is large, the $V_{\rm obs}$ and $\Vstar$
profiles are similar enough to yield good constraints on $\Udisc$. We find that these
galaxies are not all maximal discs, as their $\Udisc$ is homogeneously distributed in the
range allowed by our prior. We find the highest mass spirals ($\Mstar\gtrsim
10^{11}\Msun$) to have much better fits with a slightly larger mass-to-light ratio
($\Udisc\sim 0.7$) than the mean of our prior ($\Udisc=0.6$), consistently with previous
works who found that high mass discs are close to maximal \citep[e.g.
][]{Lapi+18,Starkman+18,Li+18}.
Smaller systems, instead, have typically a poorer inference on the mass-to-light ratio,
with about $\sim 50$ cases in which the posterior on $\Udisc$ is quite flat.
Even in these extreme cases it is nevertheless useful to let the MCMC explore the full
range of possible mass-to-light ratios ($0.01\leq\Udisc\leq 1.2$) as opposed to just fixing
a value for $\Udisc$, because this provides a more realistic estimate of the uncertainty
on the other parameters of the dark matter halo. In other words, when the inference on
$\Udisc$ is poor, it may be thought as a \emph{nuisance parameter} over which the posterior
distributions of the other two more interesting halo parameters is marginalised.

For 137 (out of 158) galaxies we obtain a unimodal posterior distribution for the halo
mass, thus we can associate a measurement and an uncertainty to $\Mhalo$; the remaining
21 galaxies have, instead, either a multi-modal or a flat posterior on the halo mass and
thus we discard them. These 21 are mostly low-mass systems ($\Mstar\lesssim 2\times
10^{9}\Msun$) and their removal does not alter in any way the high-mass end of the
population, which is the main focus of our work.
For some of the remaining 137 galaxies, we find that the NFW halo model provides
a poor fit to the observed rotation curve, as their best-fit $\chi^2$ is large.
This is not surprising, since it is well known that especially low-mass discs tend to
have slowly rising rotation curves, which makes them more compatible with having
centrally cored haloes \citep[e.g.][\citetalias{Katz+17}]{deBlok+01}.
Indeed, by re-fitting all rotation curves with a cored halo model from
\citet[][]{Burkert95}, we have found 27, mostly low-mass ($\Mstar\lesssim 10^{10}\Msun$),
systems for which such cored profile is preferred to the NFW at a 3-$\sigma$ confidence
level. For consistency we have decided to remove these 27 systems from our sample, but
in Appendix \ref{appendix} we demonstrate that their stellar and halo masses, derived by
extrapolating the Burkert profile to the virial radius, are perfectly consistent with the
picture that we present below.

%------ FIG. 1 --------------------------------------------------------------------------------
\begin{figure}
\includegraphics[width=0.5\textwidth]{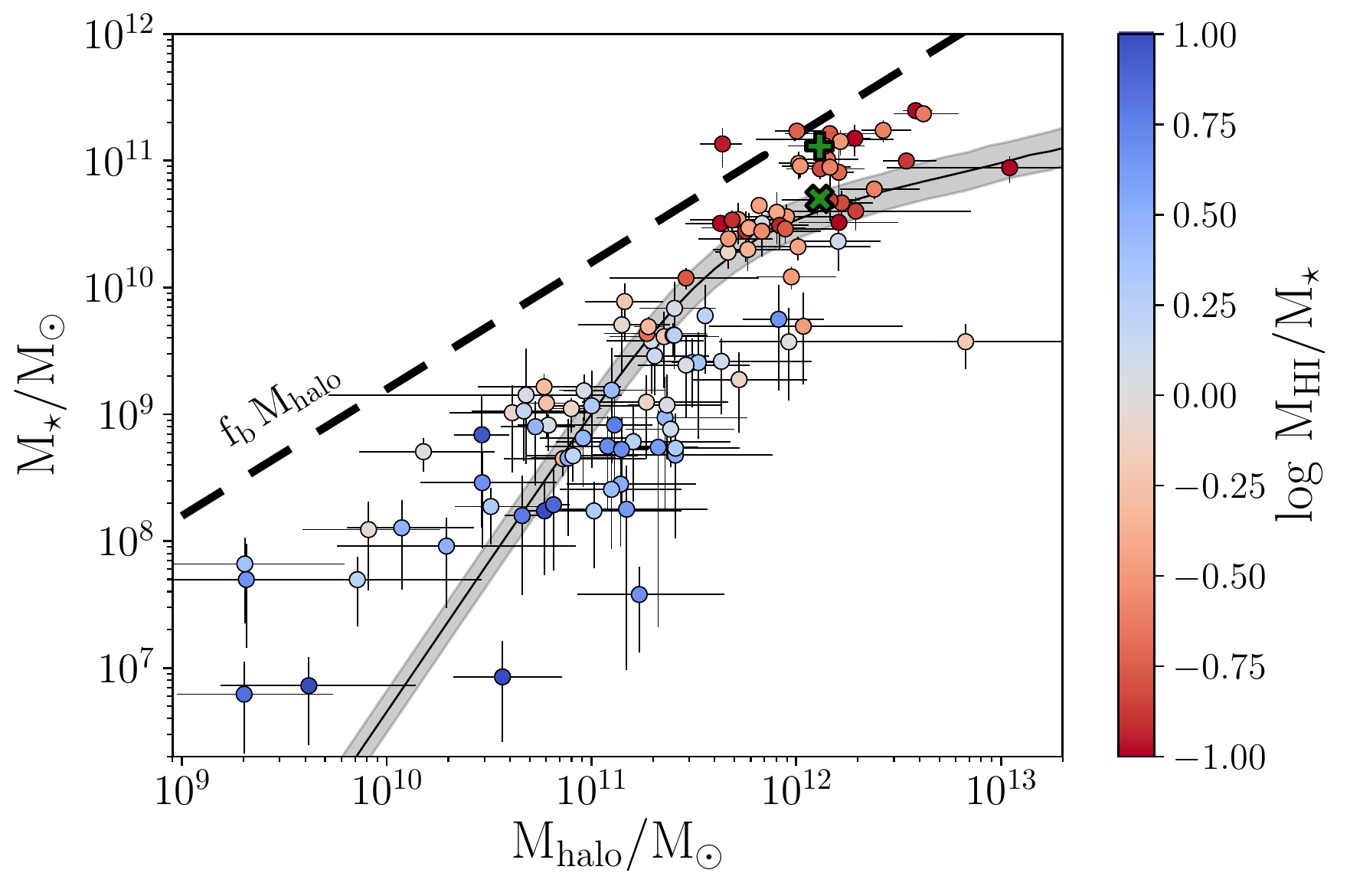}
\caption{Stellar-to-halo mass relation for 110 galaxies in the SPARC sample. The points are
         colour coded by the ratio of \hi-to-stellar mass. The stellar-to-halo mass relation
         estimated by \cite{Moster+13} using abundance matching is shown as a black dashed
         curve (with grey area representing the scatter of the relation). Galaxies that
         have converted all the available baryons in the halo into stars would lie on the
         long dashed line, whose thickness encompasses uncertainties on $\fb$.
         For reference, we also show the location of the Milky Way (cross) and
         of the Andromeda galaxy (plus) on the plot, as given by the modelling by
         \citet{PostiHelmi19} and \citet{Corbelli+10}, respectively.}
\label{fig:mstar_mhalo}
\end{figure}

%------ FIG. 2 --------------------------------------------------------------------------------
\begin{figure*}
\includegraphics[width=\textwidth]{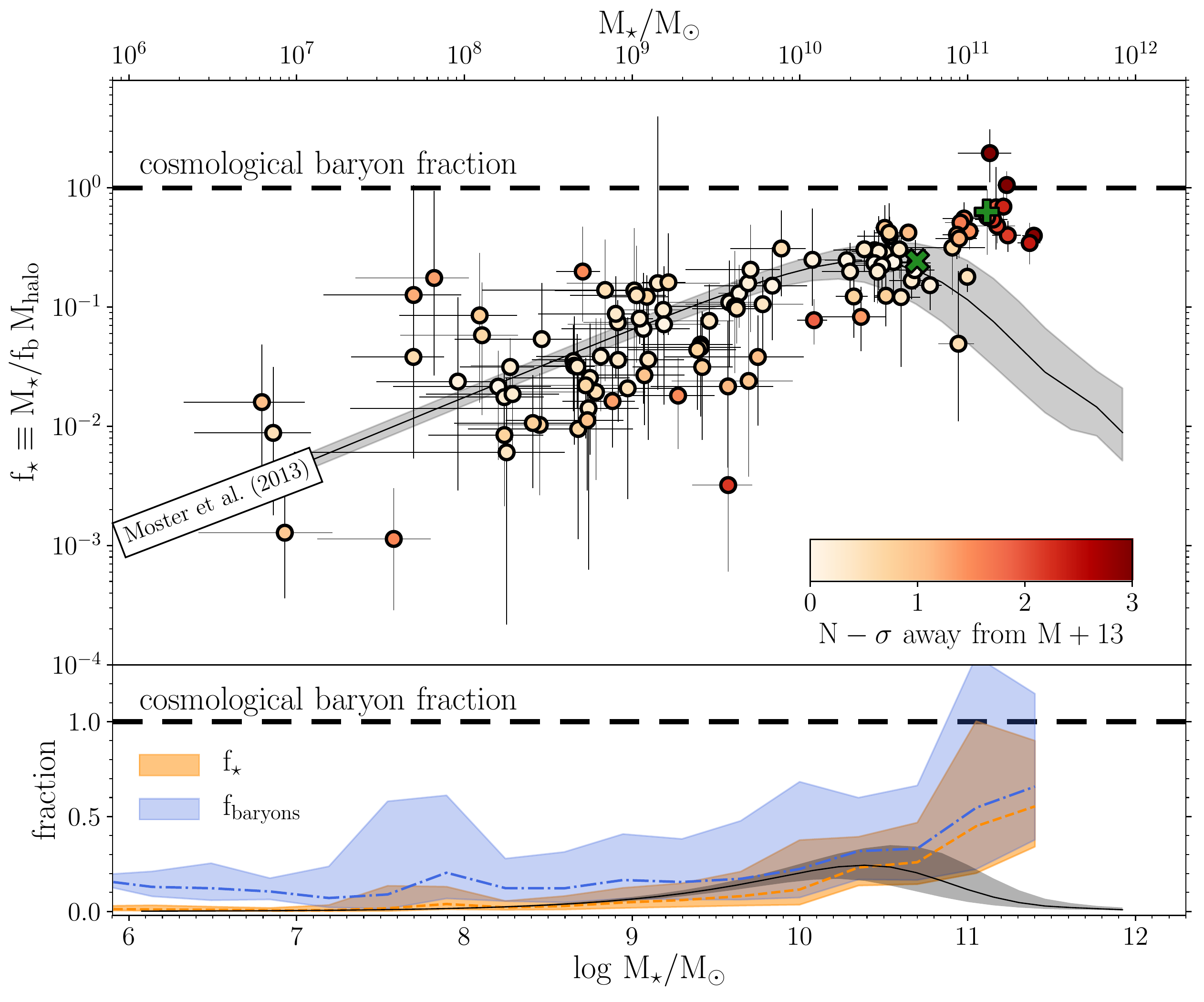}
\caption{Stellar fraction as a function of stellar mass for 110 galaxies in the SPARC sample.
         In the top panel, we show (in log-scale) the individual measurements with their
         uncertainties; in the bottom panel, we plot (in linear-scale) $\fstar$
         (orange dashed line) and $\fbaryons = \fstar + 1.4\fhi+\fhtwo$ (blue dot-dashed line,
         see text for details) in bins of $\log\,\Mstar$ (shaded areas are the $1\sigma$
         uncertainties). In both panels, the stellar-to-halo mass relation estimated by
         \cite{Moster+13} using abundance matching is shown as a black curve, with
         a shaded area representing its scatter.
         Points in the top panel are colour coded by how many standard deviations away
         the galaxy is from the \cite{Moster+13} relation, i.e. $|\fstar-f_{\star,\rm M+13}|/
         (\sigma_{\fstar}^2+\sigma_{\rm M+13}^2)^{1/2}$, where $\sigma_{\fstar}$ is the
         observed uncertainty on $\fstar$, $f_{\star,\rm M+13}$ is the value predicted by
         the abundance matching model and $\sigma_{\rm M+13}$ is the scatter of the
         \cite{Moster+13} relation.
         In both panels, galaxies that have converted all the available baryons in the halo
         into stars would lie on the long dashed line, whose thickness encompasses uncertainties
         on $\fb$. As in Fig.~\ref{fig:mstar_mhalo}, we also show the location
         of the Milky Way (cross) and of the Andromeda galaxy (plus) on the plot, as
         given by the modelling by \citet{PostiHelmi19} and \citet{Corbelli+10}, respectively.}
\label{fig:mstar_fstar}
\end{figure*}

In Figure \ref{fig:mstar_mhalo} we plot the $\Mstar-\Mhalo$ relation
for the 110 SPARC galaxies in our final sample. Points are the median of the posterior
distributions of $\Mhalo$ and $\Mstar$; the 16th-84th percentiles of the $\Mhalo$
distribution define the errorbar, while the uncertainty on the stellar mass is calculated
as in \citet[][their Eq. 5]{Lelli+16a} where the uncertainty on $\Udisc$ is given by the
16th-84th percentiles of its posterior. For comparison we also plot the $\Mstar-\Mhalo$
relation estimated by \cite{Moster+13} using abundance matching.
In general we find that the abundance matching model is in good agreement with our
measurements for $\Mstar\lesssim 5\times 10^{10}\Msun$, albeit our points have a large
scatter especially at the lowest masses. The agreement is instead much poorer at high
stellar masses, where the \citet{Moster+13} model predicts significantly larger halo masses
with respect to our estimates. Our measurements indicate that \emph{there is no sign of a
break} in the stellar-to-halo mass relation of spirals and that it is consistent with being an
increasing function of mass with roughly the same slope at all masses.
% Our results are also in very good agreement with what found
% by \citetalias{Katz+17} and sit very well in the middle of their case with uniform priors
% (their Fig. 3) and that in which they impose a prior following the \citet{Moster+13}
% $\Mstar-\Mhalo$ relation (their Fig. 5).

The tension at the high-mass end between our measurements and the abundance matching model
is much clearer if one plots the stellar fraction, i.e. $\fstar\equiv\Mstar/\fb\Mhalo$, also
sometimes called \emph{star-formation efficiency}, as a function of the stellar mass:
we show this in Figure \ref{fig:mstar_fstar}.
This plot highlights the two main findings of our work, the first being that $\fstar$
appears to increase monotonically with galaxy stellar mass \emph{with no indication of
a peak} in the range $10\leq\log\,\Mstar/\Msun\leq 11$, where most abundance matching
models find a maximum star-formation efficiency.
For instance, a galaxy with $\Mstar=2\times 10^{11}\Msun$ has $\fstar\simeq 0.04$ in the
\cite{Moster+13} model, while we find $\fstar\simeq 0.5$.
By computing the difference between the measured $\fstar$ and that expected in the
\cite{Moster+13} model, normalised by the sum in quadrature of the measured uncertainty
on $\fstar$ and of the intrinsic scatter of the model, we find the measurement for the
high-mass systems to be inconsistent at $2-3\sigma$ with the model (see the colours of
the points in Fig.~\ref{fig:mstar_fstar}).
Such a discrepancy is very robust and holds for all the tests we have run (we show in
Appendix~\ref{appendix}, Figure~\ref{fig:tests}, the $\fstar-\Mstar$ diagram in all
these cases):
\begin{itemize}
    \item we have fitted the rotation curves assuming a cored \citep{Burkert95} instead of a
          cuspy (NFW) profile. In general, this yields better fits for many low-mass
          systems, slightly larger stellar masses and smaller halo virial masses for
          all galaxies;
    \item we have used the fits recently obtained by \cite{Ghari+19}, who used
          \cite{Einasto65} halo profiles \citep[and distances and mass-to-light
          ratios from][]{Li+18}. In general, we typically find slightly smaller halo
          virial masses, but broadly consistent with our estimates with NFW profiles;
    \item we have fixed the mass-to-light ratio of the bulge and disc components to
          reasonable values suggested by stellar population synthesis models
          \citep[$\Udisc=0.5,\Ubulge=0.7$, see e.g.][]{Meidt+14,
          SchombertMcGaugh14};
    \item we tried allowing both $\Udisc$ and $\Ubulge$ to vary in our fits, with the
          additional constraint ($\Udisc\leq\Ubulge$). This has
          an effect only on the 28 (out of 110) galaxies in our final sample that have
          non-negligible bulges. We find the resulting uncertainties on $\Udisc$ to be
          typically significantly larger in this case, but never dramatic.
\end{itemize}
In all these cases the final result is that the $\fstar-\Mstar$ diagram is not
significantly different from the one presented in Fig.~\ref{fig:mstar_fstar}.

Additionally, as shown by \citet[][see their Fig. 20 and 23]{Katz+14}, the effect of
adiabatic contraction of the dark matter halos due to the formation of stellar discs has
a negligible impact on $f_\star$ for galaxies in the interested mass regime.

The other main finding highlighted by Fig.~\ref{fig:mstar_fstar} is even more
surprising: we find that all spirals with $\Mstar\gtrsim10^{11}\Msun$ have stellar
fraction very close to unity, in the range $\fstar\approx 0.3-1$, with a handful of
them being consistent with $\fstar=1$ within the uncertainties.
This implies that these galaxies were extremely
efficient at turning gas into stars and that the amount of mass collapsed in stars is a
considerable portion of the total amount of baryons expected to be associated with their
haloes. In fact, if we include also the contribution of atomic and molecular
hydrogen \citep[the latter estimated through the $M_{\rm HI}-M_{H_2}$ relation given by][]
{Catinella+18}, spirals with $\Mstar \geq 10^{11}\Msun$ are found to be consistent with
a cold baryon budget of $f_{\rm baryons}=\fstar+1.4\fhi+\fhtwo\approx 1$ within the
uncertainties \citep[where the factor 1.4 accounts for helium, e.g.][]{SPARC}, with a
mean value of $\sim 0.6$ and uncertainties of $[-0.3, +0.5]$.
Moreover, considering that galaxies are known to be surrounded by massive, hot coronae,
which are detected both in X-ray and with the Sunyaev-Zeldovich effect and account for
about $0.1-0.3\fb\Mhalo$ \citep[typically estimated statistically by stacking over many
galaxies with a given stellar mass, e.g.][and references therein]{Planck13XI,Bregman+18},
the total (cold+hot) baryon budget is easily compatible with unity at the high-mass end,
with very little room for other baryonic components.
In other words, we have found  that the most massive, regularly rotating spirals in the
nearby Universe have virtually \emph{no missing baryons}.

\section{Discussion}\label{sec:discussion}

Our analysis allowed us to have a robust and unbiased estimate of the halo virial mass
for a sample of 108 spiral galaxies in the nearby Universe using their high-quality
\hi~rotation curves. While we find good agreement with previous determinations
of the stellar-to-halo mass relation for galaxies roughly up to the mass of Milky Way
($\Mstar=5\times 10^{10}\Msun$), we also find systematically smaller halo masses (factor
$\sim 10$), corresponding to higher stellar-to-halo mass ratios, for the most massive
spirals with respect to expectations from most up-to-date abundance matching models
\citep[e.g.][]{WechslerTinker18}.

A possible explanation for this discrepancy is that, while the high-mass end
($\Mstar\gtrsim10^{11}\Msun$) of the galaxy stellar mass function is vastly dominated
by passive early-type galaxies which occupy massive ($\Mhalo\gtrsim 5\times 10^{12}\Msun$)
dark matter halos, there still exists a population of star-forming spirals that
inhabit halos of lower masses. The presence of this second population -- which is not
well represented by current abundance matching models -- implies the existence of
different evolutionary pathways for building galaxies of a given stellar mass.
This suggests that e.g. a massive system that has evolved in isolation may have had the
chance to sustain star-formation unimpeded for its entire life, potentially converting most
of its available baryons into stars. While this is certainly not the case for high-mass
early-types galaxies, which tend to live in high-density environments, it may well be
the pathway taken by the high-mass population of spirals studied in this work.
In fact, also \cite{McGaugh+10} by simply analysing the Tully-Fisher relation of
a similar sample of spirals concluded that $\fstar$ does not turn over at the highest
masses.

A discrepancy between the expected halo mass for a typical passive (red) $10^{11}\Msun$
galaxy and an active (blue) one of the same $\Mstar$, was also noted by other authors using
various probes, such as satellite kinematics \citep[e.g.][]{Conroy+07,More+11,WojtakMamon13},
galaxy-galaxy weak lensing \citep[e.g.][]{Mandelbaum+06,Mandelbaum+16,Reyes+12},
abundance matching \citep[e.g.][]{Rodriguez-Puebla+15} or combinations
\citep[e.g.][]{Dutton+10}.
The works most similar to ours are those of \citetalias{Katz+17} and \cite{Lapi+18}.
We use the same galaxy sample as in \citetalias{Katz+17} (SPARC) and we perform a similar
analysis as them, but with the crucial difference that we do not impose a prior
on halo mass that follows an $\Mstar-\Mhalo$ relation from abundance matching, which
slightly biases towards higher halo masses some of the high-mass galaxies\footnote{
Taking into account this difference in the priors used, our results are very well
compatible with theirs: our conclusions sit in the middle between their case
with uniform priors (their Fig. 3) and that in which they impose a prior following the
\citet{Moster+13} $\Mstar-\Mhalo$ relation (their Fig. 5)}.
\cite{Lapi+18}, on the other hand, have a much larger sample of spirals than ours,
but they rely on ``stacked'' rotation curves for their mass decompositions -- i.e.
they stack individual curves of galaxies in bins of absolute magnitude -- whereas
we focus on individual, well studied systems.
Finally, we notice that, amongst the detailed studies of individual systems,
i) \cite{Corbelli+10} measured the dynamical mass of M31 by decomposing its \hi~rotation
cureve, to find a surprisingly high $\fstar\simeq 0.6$, and ii) \cite{Martinsson+13}
decomposed the \hi~rotation curves of a small sample of 30 spirals from the DiskMass
Survey, to find the highest star-formation efficiencies $\fstar\gtrsim 0.3$ for their
three most massive galaxies ($\log\Mstar/\Msun\gtrsim 10.9$).
While our results align with these previous works, to our knowledge we are
the first to focus specifically on the $\fstar-\Mstar$ relation and to highlight the
fact that i) the highest-mass spirals are the most efficient galaxies at turning gas
into stars, ii) that $\fstar$ increases monotonically with stellar mass for regularly
rotating nearby discs and that iii) virtually all high-mass discs have $\gtrsim 30\%$
of the total baryons within their haloes in stars.

Our analysis establishes that the most efficient galaxies at forming stars are not
$L^\ast$ galaxies, as previously thought \citep[e.g.][]{WechslerTinker18}, but much
more massive systems, some of the most massive spiral galaxies in the nearby Universe
($\Mstar\gtrsim 10^{11}\Msun$). Not only the galactic star-formation efficiency peaks at
much larger masses than we knew before, but we also showed that \emph{several
massive discs have efficiencies $\fstar$ of the order unity}. This result alone is
of key importance since it demonstrates that there is no universal physical mechanism
that sets the maximum star-formation efficiency to $20-30\%$.

Furthermore, the fact that some massive galaxies with high $\fstar$ exist has fundamental
implications for star-formation quenching. Since these galaxies live in haloes with
$\Mhalo\sim 2-5\times 10^{12}\Msun$, if mass is the main driver of quenching and if a
critical mass for quenching exists \citep[e.g. as expected in scenarios where virial
shock heating of the circumgalactic medium is the key process, see][]{BirnboimDekel03,
DekelBirnboim08}, then it follows that this critical mass can not be smaller
than $\sim 5\times 10^{12}\Msun$, which is almost an order of magnitude higher than
previously thought \citep[e.g.][]{DekelBirnboim06}. Interestingly, such a high threshold
is instead expected in scenarios where the accretion of cool gas is hampered
(``starvation''), e.g. by the high virial temperature of the circumgalactic gas in a
galactic fountain cycle \citep[e.g.][]{Armillotta+16} or by the complex interplay of
radiative cooling and feedback in the smooth gas accretion from cold filaments
\citep[e.g.][]{vandeVoort+11}.

Even if we have measured high $\fstar$ for some massive spirals, still the vast majority
of galaxies living in $\Mhalo>10^{12}\Msun$ haloes has $\fstar\ll 1$, which means that
they managed to efficiently quench their star-formation.
Our results imply that since mass can not be the major player in quenching galaxies, at
least for $\Mhalo \lesssim 5\times 10^{12}\Msun$, and some other mechanism must
play a fundamental role in the transition from actively to passively star-forming. One
of the main suspects is clearly environment, since gas removal happens more frequently
and also gas accretion is more difficult in high-density environments
\citep[e.g.][]{Peng+10,vandeVoort+17}. Another is the powerful feedback from the active
galactic nucleus (AGN), which can episodically suppress any gas condensation throughout
the galaxy \citep[e.g.][]{Croton+06,Fabian12}. Finally, another key process is the
interaction with other galaxies, with passive galaxies being hosted in haloes with an
active merger history, which can result in bursty star-formation histories and
subsequent suppressive stellar/AGN feedback \citep[e.g.][]{Cox+06b,Gabor+10}. This latter
scenario also naturally accounts for the morphological transformation of disc galaxies,
living in haloes with quiet merger histories, to spheroids, which are the dominant galaxy
population at the high-mass end, where also mergers are more frequent
\citep[e.g.][]{Cox+06a}.
% To test this case, one could envision revisiting current abundance
% matching models by making them dependent on two parameters, mass and merger history:
% the stellar mass function of the blue (red) galaxy population should be matched
% with the halo mass function of haloes with quiet (violent) merger histories.
This scenario is, in principle, testable both with current cosmological simulations and
with a new abundance matching model which depends also on secondary halo parameters,
such as merger history or formation time, and it is able to predict not only stellar
masses but also other galaxy properties, such as morphology or colour.

\section*{Acknowledgements}
We thank E. Corbelli, B. Famaey, A. Lapi, F. Lelli, A. Robertson, J. Sellwood and
F. van den Bosch for useful discussions and A. Ghari for making their Einasto fits
available to us. LP acknowledges financial support from a VICI grant from the
Netherlands Organisation for Scientific Research (NWO) and from the Centre National
d’Etudes Spatiales (CNES).

%-------------------------------------------------------------------

\bibliographystyle{aa} % style aa.bst
\bibliography{refs} % your references Yourfile.bib

\begin{thebibliography}{68}
\expandafter\ifx\csname natexlab\endcsname\relax\def\natexlab#1{#1}\fi

\bibitem[{{Armillotta} {et~al.}(2016){Armillotta}, {Fraternali}, \&
  {Marinacci}}]{Armillotta+16}
{Armillotta}, L., {Fraternali}, F., \& {Marinacci}, F. 2016, \mnras, 462, 4157

\bibitem[{{Behroozi} {et~al.}(2010){Behroozi}, {Conroy}, \&
  {Wechsler}}]{Behroozi+10}
{Behroozi}, P.~S., {Conroy}, C., \& {Wechsler}, R.~H. 2010, \apj, 717, 379

\bibitem[{{Birnboim} \& {Dekel}(2003)}]{BirnboimDekel03}
{Birnboim}, Y. \& {Dekel}, A. 2003, \mnras, 345, 349

\bibitem[{{Bregman}(2007)}]{Bregman07}
{Bregman}, J.~N. 2007, \araa, 45, 221

\bibitem[{{Bregman} {et~al.}(2018){Bregman}, {Anderson}, {Miller},
  {Hodges-Kluck}, {Dai}, {Li}, {Li}, \& {Qu}}]{Bregman+18}
{Bregman}, J.~N., {Anderson}, M.~E., {Miller}, M.~J., {et~al.} 2018, \apj, 862,
  3

\bibitem[{{Burkert}(1995)}]{Burkert95}
{Burkert}, A. 1995, \apjl, 447, L25

\bibitem[{{Cappellari} {et~al.}(2013){Cappellari}, {Scott}, {Alatalo}, {Blitz},
  {Bois}, {Bournaud}, {Bureau}, {Crocker}, {Davies}, {Davis}, {de Zeeuw},
  {Duc}, {Emsellem}, {Khochfar}, {Krajnovi{\'c}}, {Kuntschner}, {McDermid},
  {Morganti}, {Naab}, {Oosterloo}, {Sarzi}, {Serra}, {Weijmans}, \&
  {Young}}]{Cappellari+13}
{Cappellari}, M., {Scott}, N., {Alatalo}, K., {et~al.} 2013, \mnras, 432, 1709

\bibitem[{{Catinella} {et~al.}(2018){Catinella}, {Saintonge}, {Janowiecki},
  {Cortese}, {Dav{\'e}}, {Lemonias}, {Cooper}, {Schiminovich}, {Hummels},
  {Fabello}, {Ger{\'e}b}, {Kilborn}, \& {Wang}}]{Catinella+18}
{Catinella}, B., {Saintonge}, A., {Janowiecki}, S., {et~al.} 2018, \mnras, 476,
  875

\bibitem[{{Conroy} {et~al.}(2007){Conroy}, {Prada}, {Newman}, {Croton}, {Coil},
  {Conselice}, {Cooper}, {Davis}, {Faber}, {Gerke}, {Guhathakurta}, {Klypin},
  {Koo}, \& {Yan}}]{Conroy+07}
{Conroy}, C., {Prada}, F., {Newman}, J.~A., {et~al.} 2007, \apj, 654, 153

\bibitem[{{Corbelli} {et~al.}(2010){Corbelli}, {Lorenzoni}, {Walterbos},
  {Braun}, \& {Thilker}}]{Corbelli+10}
{Corbelli}, E., {Lorenzoni}, S., {Walterbos}, R., {Braun}, R., \& {Thilker}, D.
  2010, \aap, 511, A89

\bibitem[{{Cox} {et~al.}(2006{\natexlab{a}}){Cox}, {Dutta}, {Di Matteo},
  {Hernquist}, {Hopkins}, {Robertson}, \& {Springel}}]{Cox+06a}
{Cox}, T.~J., {Dutta}, S.~N., {Di Matteo}, T., {et~al.} 2006{\natexlab{a}},
  \apj, 650, 791

\bibitem[{{Cox} {et~al.}(2006{\natexlab{b}}){Cox}, {Jonsson}, {Primack}, \&
  {Somerville}}]{Cox+06b}
{Cox}, T.~J., {Jonsson}, P., {Primack}, J.~R., \& {Somerville}, R.~S.
  2006{\natexlab{b}}, \mnras, 373, 1013

\bibitem[{{Croton} {et~al.}(2006){Croton}, {Springel}, {White}, {De Lucia},
  {Frenk}, {Gao}, {Jenkins}, {Kauffmann}, {Navarro}, \& {Yoshida}}]{Croton+06}
{Croton}, D.~J., {Springel}, V., {White}, S.~D.~M., {et~al.} 2006, \mnras, 365,
  11

\bibitem[{{de Blok} {et~al.}(2001){de Blok}, {McGaugh}, {Bosma}, \&
  {Rubin}}]{deBlok+01}
{de Blok}, W.~J.~G., {McGaugh}, S.~S., {Bosma}, A., \& {Rubin}, V.~C. 2001,
  \apjl, 552, L23

\bibitem[{{Dekel} \& {Birnboim}(2006)}]{DekelBirnboim06}
{Dekel}, A. \& {Birnboim}, Y. 2006, \mnras, 368, 2

\bibitem[{{Dekel} \& {Birnboim}(2008)}]{DekelBirnboim08}
{Dekel}, A. \& {Birnboim}, Y. 2008, \mnras, 383, 119

\bibitem[{{Desmond} \& {Wechsler}(2015)}]{DesmondWechsler15}
{Desmond}, H. \& {Wechsler}, R.~H. 2015, \mnras, 454, 322

\bibitem[{{Dutton} {et~al.}(2010){Dutton}, {Conroy}, {van den Bosch}, {Prada},
  \& {More}}]{Dutton+10}
{Dutton}, A.~A., {Conroy}, C., {van den Bosch}, F.~C., {Prada}, F., \& {More},
  S. 2010, \mnras, 407, 2

\bibitem[{{Dutton} \& {Macci{\`o}}(2014)}]{DuttonMaccio14}
{Dutton}, A.~A. \& {Macci{\`o}}, A.~V. 2014, \mnras, 441, 3359

\bibitem[{{Einasto}(1965)}]{Einasto65}
{Einasto}, J. 1965, Trudy Astrofizicheskogo Instituta Alma-Ata, 5, 87

\bibitem[{{Fabian}(2012)}]{Fabian12}
{Fabian}, A.~C. 2012, \araa, 50, 455

\bibitem[{{Foreman-Mackey} {et~al.}(2013){Foreman-Mackey}, {Hogg}, {Lang}, \&
  {Goodman}}]{emcee}
{Foreman-Mackey}, D., {Hogg}, D.~W., {Lang}, D., \& {Goodman}, J. 2013, \pasp,
  125, 306

\bibitem[{{Fukugita} {et~al.}(1998){Fukugita}, {Hogan}, \&
  {Peebles}}]{Fukugita+98}
{Fukugita}, M., {Hogan}, C.~J., \& {Peebles}, P.~J.~E. 1998, \apj, 503, 518

\bibitem[{{Gabor} {et~al.}(2010){Gabor}, {Dav{\'e}}, {Finlator}, \&
  {Oppenheimer}}]{Gabor+10}
{Gabor}, J.~M., {Dav{\'e}}, R., {Finlator}, K., \& {Oppenheimer}, B.~D. 2010,
  \mnras, 407, 749

\bibitem[{{Ghari} {et~al.}(2018){Ghari}, {Famaey}, {Laporte}, \&
  {Haghi}}]{Ghari+19}
{Ghari}, A., {Famaey}, B., {Laporte}, C., \& {Haghi}, H. 2018, arXiv e-prints
  [\eprint[arXiv]{1811.06554}]

\bibitem[{{Katz} {et~al.}(2017){Katz}, {Lelli}, {McGaugh}, {Di Cintio},
  {Brook}, \& {Schombert}}]{Katz+17}
{Katz}, H., {Lelli}, F., {McGaugh}, S.~S., {et~al.} 2017, \mnras, 466, 1648

\bibitem[{{Katz} {et~al.}(2014){Katz}, {McGaugh}, {Sellwood}, \& {de
  Blok}}]{Katz+14}
{Katz}, H., {McGaugh}, S.~S., {Sellwood}, J.~A., \& {de Blok}, W.~J.~G. 2014,
  \mnras, 439, 1897

\bibitem[{{Kelvin} {et~al.}(2014){Kelvin}, {Driver}, {Robotham}, {Taylor},
  {Graham}, {Alpaslan}, {Baldry}, {Bamford}, {Bauer}, {Bland-Hawthorn},
  {Brown}, {Colless}, {Conselice}, {Holwerda}, {Hopkins}, {Lara-L{\'o}pez},
  {Liske}, {L{\'o}pez-S{\'a}nchez}, {Loveday}, {Norberg}, {Phillipps},
  {Popescu}, {Prescott}, {Sansom}, \& {Tuffs}}]{Kelvin+14}
{Kelvin}, L.~S., {Driver}, S.~P., {Robotham}, A.~S.~G., {et~al.} 2014, \mnras,
  444, 1647

\bibitem[{{Kravtsov} {et~al.}(2004){Kravtsov}, {Berlind}, {Wechsler}, {Klypin},
  {Gottl{\"o}ber}, {Allgood}, \& {Primack}}]{Kravtsov+04}
{Kravtsov}, A.~V., {Berlind}, A.~A., {Wechsler}, R.~H., {et~al.} 2004, \apj,
  609, 35

\bibitem[{{Lange} {et~al.}(2018){Lange}, {van den Bosch}, {Zentner}, {Wang}, \&
  {Villarreal}}]{Lange+18}
{Lange}, J.~U., {van den Bosch}, F.~C., {Zentner}, A.~R., {Wang}, K., \&
  {Villarreal}, A.~S. 2018, ArXiv e-prints [\eprint[arXiv]{1811.03596}]

\bibitem[{{Lapi} {et~al.}(2018){Lapi}, {Salucci}, \& {Danese}}]{Lapi+18}
{Lapi}, A., {Salucci}, P., \& {Danese}, L. 2018, \apj, 859, 2

\bibitem[{{Leauthaud} {et~al.}(2012){Leauthaud}, {Tinker}, {Bundy}, {Behroozi},
  {Massey}, {Rhodes}, {George}, {Kneib}, {Benson}, {Wechsler}, {Busha},
  {Capak}, {Cort{\^e}s}, {Ilbert}, {Koekemoer}, {Le F{\`e}vre}, {Lilly},
  {McCracken}, {Salvato}, {Schrabback}, {Scoville}, {Smith}, \&
  {Taylor}}]{Leauthaud+12}
{Leauthaud}, A., {Tinker}, J., {Bundy}, K., {et~al.} 2012, \apj, 744, 159

\bibitem[{{Lelli} {et~al.}(2016{\natexlab{a}}){Lelli}, {McGaugh}, \&
  {Schombert}}]{SPARC}
{Lelli}, F., {McGaugh}, S.~S., \& {Schombert}, J.~M. 2016{\natexlab{a}}, \aj,
  152, 157

\bibitem[{{Lelli} {et~al.}(2016{\natexlab{b}}){Lelli}, {McGaugh}, \&
  {Schombert}}]{Lelli+16a}
{Lelli}, F., {McGaugh}, S.~S., \& {Schombert}, J.~M. 2016{\natexlab{b}}, \apjl,
  816, L14

\bibitem[{{Li} {et~al.}(2018){Li}, {Lelli}, {McGaugh}, \& {Schombert}}]{Li+18}
{Li}, P., {Lelli}, F., {McGaugh}, S., \& {Schombert}, J. 2018, \aap, 615, A3

\bibitem[{{Mandelbaum} {et~al.}(2006){Mandelbaum}, {Seljak}, {Kauffmann},
  {Hirata}, \& {Brinkmann}}]{Mandelbaum+06}
{Mandelbaum}, R., {Seljak}, U., {Kauffmann}, G., {Hirata}, C.~M., \&
  {Brinkmann}, J. 2006, \mnras, 368, 715

\bibitem[{{Mandelbaum} {et~al.}(2016){Mandelbaum}, {Wang}, {Zu}, {White},
  {Henriques}, \& {More}}]{Mandelbaum+16}
{Mandelbaum}, R., {Wang}, W., {Zu}, Y., {et~al.} 2016, \mnras, 457, 3200

\bibitem[{{Martinsson} {et~al.}(2013){Martinsson}, {Verheijen}, {Westfall},
  {Bershady}, {Andersen}, \& {Swaters}}]{Martinsson+13}
{Martinsson}, T.~P.~K., {Verheijen}, M.~A.~W., {Westfall}, K.~B., {et~al.}
  2013, \aap, 557, A131

\bibitem[{{McConnachie}(2012)}]{McConnachie12}
{McConnachie}, A.~W. 2012, \aj, 144, 4

\bibitem[{{McGaugh} \& {Schombert}(2014)}]{McGaughSchombert14}
{McGaugh}, S.~S. \& {Schombert}, J.~M. 2014, \aj, 148, 77

\bibitem[{{McGaugh} {et~al.}(2010){McGaugh}, {Schombert}, {de Blok}, \&
  {Zagursky}}]{McGaugh+10}
{McGaugh}, S.~S., {Schombert}, J.~M., {de Blok}, W.~J.~G., \& {Zagursky}, M.~J.
  2010, \apjl, 708, L14

\bibitem[{{Meidt} {et~al.}(2014){Meidt}, {Schinnerer}, {van de Ven},
  {Zaritsky}, {Peletier}, {Knapen}, {Sheth}, {Regan}, {Querejeta},
  {Mu{\~n}oz-Mateos}, {Kim}, {Hinz}, {Gil de Paz}, {Athanassoula}, {Bosma},
  {Buta}, {Cisternas}, {Ho}, {Holwerda}, {Skibba}, {Laurikainen}, {Salo},
  {Gadotti}, {Laine}, {Erroz-Ferrer}, {Comer{\'o}n}, {Men{\'e}ndez-Delmestre},
  {Seibert}, \& {Mizusawa}}]{Meidt+14}
{Meidt}, S.~E., {Schinnerer}, E., {van de Ven}, G., {et~al.} 2014, \apj, 788,
  144

\bibitem[{{More} {et~al.}(2011){More}, {van den Bosch}, {Cacciato}, {Skibba},
  {Mo}, \& {Yang}}]{More+11}
{More}, S., {van den Bosch}, F.~C., {Cacciato}, M., {et~al.} 2011, \mnras, 410,
  210

\bibitem[{{Moster} {et~al.}(2013){Moster}, {Naab}, \& {White}}]{Moster+13}
{Moster}, B.~P., {Naab}, T., \& {White}, S.~D.~M. 2013, \mnras, 428, 3121

\bibitem[{{Navarro} {et~al.}(1996){Navarro}, {Frenk}, \& {White}}]{NFW}
{Navarro}, J.~F., {Frenk}, C.~S., \& {White}, S.~D.~M. 1996, \apj, 462, 563

\bibitem[{{Papastergis} {et~al.}(2012){Papastergis}, {Cattaneo}, {Huang},
  {Giovanelli}, \& {Haynes}}]{Papastergis+12}
{Papastergis}, E., {Cattaneo}, A., {Huang}, S., {Giovanelli}, R., \& {Haynes},
  M.~P. 2012, \apj, 759, 138

\bibitem[{{Peng} {et~al.}(2010){Peng}, {Lilly}, {Kova{\v c}}, {Bolzonella},
  {Pozzetti}, {Renzini}, {Zamorani}, {Ilbert}, {Knobel}, {Iovino}, {Maier},
  {Cucciati}, {Tasca}, {Carollo}, {Silverman}, {Kampczyk}, {de Ravel},
  {Sanders}, {Scoville}, {Contini}, {Mainieri}, {Scodeggio}, {Kneib}, {Le
  F{\`e}vre}, {Bardelli}, {Bongiorno}, {Caputi}, {Coppa}, {de la Torre},
  {Franzetti}, {Garilli}, {Lamareille}, {Le Borgne}, {Le Brun}, {Mignoli},
  {Perez Montero}, {Pello}, {Ricciardelli}, {Tanaka}, {Tresse}, {Vergani},
  {Welikala}, {Zucca}, {Oesch}, {Abbas}, {Barnes}, {Bordoloi}, {Bottini},
  {Cappi}, {Cassata}, {Cimatti}, {Fumana}, {Hasinger}, {Koekemoer},
  {Leauthaud}, {Maccagni}, {Marinoni}, {McCracken}, {Memeo}, {Meneux}, {Nair},
  {Porciani}, {Presotto}, \& {Scaramella}}]{Peng+10}
{Peng}, Y.-j., {Lilly}, S.~J., {Kova{\v c}}, K., {et~al.} 2010, \apj, 721, 193

\bibitem[{{Persic} \& {Salucci}(1992)}]{PersicSalucci92}
{Persic}, M. \& {Salucci}, P. 1992, \mnras, 258, 14P

\bibitem[{{Persic} {et~al.}(1996){Persic}, {Salucci}, \& {Stel}}]{Persic+96}
{Persic}, M., {Salucci}, P., \& {Stel}, F. 1996, \mnras, 281, 27

\bibitem[{{Planck Collaboration} {et~al.}(2013){Planck Collaboration}, {Ade},
  {Aghanim}, {Arnaud}, {Ashdown}, {Atrio-Barandela}, {Aumont}, {Baccigalupi},
  {Balbi}, {Banday}, \& et~al.}]{Planck13XI}
{Planck Collaboration}, {Ade}, P.~A.~R., {Aghanim}, N., {et~al.} 2013, \aap,
  557, A52

\bibitem[{{Planck Collaboration} {et~al.}(2018){Planck Collaboration},
  {Aghanim}, {Akrami}, {Ashdown}, {Aumont}, {Baccigalupi}, {Ballardini},
  {Banday}, {Barreiro}, {Bartolo}, {Basak}, {Battye}, {Benabed}, {Bernard},
  {Bersanelli}, \& {Bielewicz}}]{Planck18}
{Planck Collaboration}, {Aghanim}, N., {Akrami}, Y., {et~al.} 2018, ArXiv
  e-prints [\eprint[arXiv]{1807.06209}]

\bibitem[{{Posti} \& {Helmi}(2019)}]{PostiHelmi19}
{Posti}, L. \& {Helmi}, A. 2019, \aap, 621, A56

\bibitem[{{Read} {et~al.}(2017){Read}, {Iorio}, {Agertz}, \&
  {Fraternali}}]{Read+17}
{Read}, J.~I., {Iorio}, G., {Agertz}, O., \& {Fraternali}, F. 2017, \mnras,
  467, 2019

\bibitem[{{Reyes} {et~al.}(2012){Reyes}, {Mandelbaum}, {Gunn}, {Nakajima},
  {Seljak}, \& {Hirata}}]{Reyes+12}
{Reyes}, R., {Mandelbaum}, R., {Gunn}, J.~E., {et~al.} 2012, \mnras, 425, 2610

\bibitem[{{Rodr{\'{\i}}guez-Puebla} {et~al.}(2015){Rodr{\'{\i}}guez-Puebla},
  {Avila-Reese}, {Yang}, {Foucaud}, {Drory}, \& {Jing}}]{Rodriguez-Puebla+15}
{Rodr{\'{\i}}guez-Puebla}, A., {Avila-Reese}, V., {Yang}, X., {et~al.} 2015,
  \apj, 799, 130

\bibitem[{{Salucci} \& {Burkert}(2000)}]{SalucciBurkert00}
{Salucci}, P. \& {Burkert}, A. 2000, \apjl, 537, L9

\bibitem[{{Schombert} \& {McGaugh}(2014)}]{SchombertMcGaugh14}
{Schombert}, J. \& {McGaugh}, S. 2014, \pasa, 31, e036

\bibitem[{{Starkman} {et~al.}(2018){Starkman}, {Lelli}, {McGaugh}, \&
  {Schombert}}]{Starkman+18}
{Starkman}, N., {Lelli}, F., {McGaugh}, S., \& {Schombert}, J. 2018, \mnras,
  480, 2292

\bibitem[{{Tumlinson} {et~al.}(2017){Tumlinson}, {Peeples}, \&
  {Werk}}]{Tumlinson+17}
{Tumlinson}, J., {Peeples}, M.~S., \& {Werk}, J.~K. 2017, \araa, 55, 389

\bibitem[{{Vale} \& {Ostriker}(2004)}]{ValeOstriker04}
{Vale}, A. \& {Ostriker}, J.~P. 2004, \mnras, 353, 189

\bibitem[{{van de Voort} {et~al.}(2017){van de Voort}, {Bah{\'e}}, {Bower},
  {Correa}, {Crain}, {Schaye}, \& {Theuns}}]{vandeVoort+17}
{van de Voort}, F., {Bah{\'e}}, Y.~M., {Bower}, R.~G., {et~al.} 2017, \mnras,
  466, 3460

\bibitem[{{van de Voort} {et~al.}(2011){van de Voort}, {Schaye}, {Booth},
  {Haas}, \& {Dalla Vecchia}}]{vandeVoort+11}
{van de Voort}, F., {Schaye}, J., {Booth}, C.~M., {Haas}, M.~R., \& {Dalla
  Vecchia}, C. 2011, \mnras, 414, 2458

\bibitem[{{van den Bosch} {et~al.}(2004){van den Bosch}, {Norberg}, {Mo}, \&
  {Yang}}]{vandenBosch+04}
{van den Bosch}, F.~C., {Norberg}, P., {Mo}, H.~J., \& {Yang}, X. 2004, \mnras,
  352, 1302

\bibitem[{{Wechsler} \& {Tinker}(2018)}]{WechslerTinker18}
{Wechsler}, R.~H. \& {Tinker}, J.~L. 2018, \araa, 56, 435

\bibitem[{{White} \& {Rees}(1978)}]{WhiteRees78}
{White}, S.~D.~M. \& {Rees}, M.~J. 1978, \mnras, 183, 341

\bibitem[{{Wojtak} \& {Mamon}(2013)}]{WojtakMamon13}
{Wojtak}, R. \& {Mamon}, G.~A. 2013, \mnras, 428, 2407

\bibitem[{{Yang} {et~al.}(2008){Yang}, {Mo}, \& {van den Bosch}}]{Yang+08}
{Yang}, X., {Mo}, H.~J., \& {van den Bosch}, F.~C. 2008, \apj, 676, 248

\bibitem[{{Zheng} {et~al.}(2007){Zheng}, {Coil}, \& {Zehavi}}]{Zheng+07}
{Zheng}, Z., {Coil}, A.~L., \& {Zehavi}, I. 2007, \apj, 667, 760

\end{thebibliography}

\newpage
\appendix
\section{Supplementary material} \label{appendix}

%------ TAB. 2 --------------------------------------------------------------------------------
\clearpage
\onecolumn
\begin{landscape}
\begin{center}
\begin{longtable}{lcccccccccccccc}
\caption{Results of the fits for individual galaxies. The near-infrared luminosity $L_{[3.6]}$
is given in solar luminosities; the posteriors of the three model parameters, disc mass-to-light
ratio $\Udisc$, halo mass $\Mhalo$ and concentration $c$ are represented with their 50th-16th-84th
percentiles; $\chi^2_{\rm red}$ is the reduced $\chi^2$ (Eq.~\ref{eq:likelihood}) for the best
fit model; the posterior on the derived parameter $\fstar=\Mstar/\fb\Mhalo$ is represented with
its 50th-16th-84th percentiles.}
\label{tab:results}
 \\
Name & $\log\,L_{[3.6]}$ & $\Udisc$ & 16th & 84th & $\log\,\Mhalo$ & 16th & 84th & $\log\,c$ & 16th & 84th & $\chi^2_{\rm red}$ & $\fstar$ & 16th & 84th  \vspace{.1cm}\\
\hline\hline \\[-.2cm]
\endfirsthead

\multicolumn{15}{c}%
{{\bfseries \tablename\ \thetable{} -- continued from previous page}} \\
Name & $\log\,L_{[3.6]}$ & $\Udisc$ & 16th & 84th & $\log\,\Mhalo$ & 16th & 84th & $\log\,c$ & 16th & 84th & $\chi^2_{\rm red}$ & $\fstar$ & 16th & 84th  \vspace{.1cm}\\
\hline\hline \\[-.2cm]
\endhead

\hline \multicolumn{15}{|r|}{{Continued on next page}} \\ \hline
\endfoot

\hline \hline
\endlastfoot

D512-2  & 8.51 & 0.62 & 0.22 & 1.02 & 9.91 & 9.59 & 10.26 & 0.98 & 0.86 & 1.11 & 1.05  & 0.0852 & 0.0157 & 0.2845  \\
DDO064  & 8.20 & 0.60 & 0.21 & 1.00 & 10.29 & 9.76 & 10.92 & 1.00 & 0.83 & 1.17 & 1.07  & 0.0237 & 0.0029 & 0.1213  \\
DDO170  & 8.73 & 0.38 & 0.12 & 0.80 & 10.66 & 10.58 & 10.76 & 0.82 & 0.74 & 0.88 & 2.73  & 0.0215 & 0.0052 & 0.0431  \\
ESO116-G012  & 9.63 & 0.44 & 0.20 & 0.69 & 11.72 & 11.49 & 12.05 & 0.89 & 0.74 & 1.01 & 2.52  & 0.0181 & 0.0064 & 0.0374  \\
ESO444-G084  & 7.85 & 0.60 & 0.21 & 0.99 & 11.23 & 10.93 & 11.65 & 0.92 & 0.79 & 1.03 & 0.76  & 0.0011 & 0.0003 & 0.0030  \\
F565-V2  & 8.75 & 0.58 & 0.19 & 1.00 & 11.14 & 10.88 & 11.51 & 0.89 & 0.73 & 1.02 & 1.10  & 0.0103 & 0.0026 & 0.0280  \\
F568-V1  & 9.58 & 0.70 & 0.27 & 1.05 & 11.63 & 11.30 & 12.07 & 1.01 & 0.84 & 1.15 & 0.30  & 0.0314 & 0.0077 & 0.0919  \\
F574-1  & 9.82 & 0.68 & 0.27 & 1.03 & 11.29 & 11.07 & 11.55 & 0.92 & 0.80 & 1.03 & 1.84  & 0.1097 & 0.0312 & 0.2466  \\
F583-1  & 8.99 & 0.57 & 0.18 & 0.98 & 11.08 & 10.77 & 11.42 & 0.86 & 0.73 & 0.98 & 2.11  & 0.0255 & 0.0058 & 0.0731  \\
F583-4  & 9.23 & 0.62 & 0.21 & 1.02 & 10.61 & 10.31 & 10.98 & 0.98 & 0.82 & 1.11 & 0.44  & 0.1377 & 0.0263 & 0.4602  \\
NGC0024  & 9.59 & 1.02 & 0.77 & 1.15 & 11.27 & 11.06 & 11.56 & 1.06 & 0.91 & 1.20 & 0.66  & 0.1306 & 0.0670 & 0.2269  \\
NGC0100  & 9.51 & 0.29 & 0.09 & 0.60 & 11.36 & 11.02 & 11.76 & 0.85 & 0.70 & 0.97 & 1.20  & 0.0208 & 0.0024 & 0.0820  \\
NGC0247  & 9.87 & 0.64 & 0.25 & 1.01 & 11.35 & 11.09 & 11.62 & 0.82 & 0.72 & 0.90 & 2.14  & 0.1013 & 0.0280 & 0.2969  \\
NGC0289  & 10.86 & 0.59 & 0.43 & 0.76 & 11.83 & 11.74 & 11.94 & 0.91 & 0.77 & 1.05 & 1.95  & 0.2833 & 0.1765 & 0.4227  \\
NGC0300  & 9.47 & 0.46 & 0.17 & 0.79 & 11.37 & 11.18 & 11.63 & 0.89 & 0.75 & 1.01 & 0.72  & 0.0268 & 0.0102 & 0.0573  \\
NGC0801  & 11.49 & 0.56 & 0.52 & 0.60 & 12.00 & 11.90 & 12.14 & 0.77 & 0.63 & 0.90 & 6.80  & 1.0564 & 0.7746 & 1.3789  \\
NGC1003  & 9.83 & 0.46 & 0.24 & 0.66 & 11.49 & 11.39 & 11.62 & 0.78 & 0.67 & 0.88 & 3.09  & 0.0485 & 0.0240 & 0.0746  \\
NGC1090  & 10.86 & 0.48 & 0.36 & 0.59 & 11.72 & 11.63 & 11.84 & 0.94 & 0.80 & 1.07 & 2.50  & 0.3931 & 0.2445 & 0.5734  \\
NGC1705  & 8.73 & 0.99 & 0.72 & 1.15 & 10.86 & 10.57 & 11.26 & 1.16 & 0.99 & 1.31 & 0.66  & 0.0352 & 0.0133 & 0.0712  \\
NGC2403  & 10.00 & 0.42 & 0.30 & 0.53 & 11.40 & 11.33 & 11.49 & 1.14 & 1.06 & 1.23 & 9.47  & 0.1012 & 0.0828 & 0.1164  \\
NGC2683  & 10.91 & 0.66 & 0.58 & 0.73 & 11.63 & 11.46 & 11.82 & 0.96 & 0.82 & 1.11 & 1.31  & 0.4620 & 0.2927 & 0.7192  \\
NGC2841  & 11.27 & 0.87 & 0.79 & 0.94 & 12.54 & 12.42 & 12.69 & 0.88 & 0.76 & 1.00 & 1.81  & 0.1796 & 0.1335 & 0.2283  \\
NGC2903  & 10.91 & 0.37 & 0.31 & 0.41 & 11.75 & 11.67 & 11.85 & 1.24 & 1.14 & 1.34 & 7.61  & 0.3001 & 0.1718 & 0.4448  \\
NGC2915  & 8.81 & 0.56 & 0.19 & 0.97 & 11.10 & 10.85 & 11.44 & 1.03 & 0.86 & 1.18 & 0.98  & 0.0106 & 0.0030 & 0.0270  \\
NGC2955  & 11.50 & 0.47 & 0.44 & 0.51 & 12.13 & 11.80 & 12.48 & 0.88 & 0.71 & 1.03 & 4.81  & 0.6863 & 0.3052 & 1.4911  \\
NGC2998  & 11.18 & 0.62 & 0.48 & 0.74 & 12.01 & 11.91 & 12.13 & 0.91 & 0.76 & 1.06 & 2.74  & 0.5532 & 0.3896 & 0.7568  \\
NGC3198  & 10.58 & 0.51 & 0.38 & 0.61 & 11.67 & 11.60 & 11.75 & 0.98 & 0.87 & 1.09 & 1.43  & 0.2475 & 0.1981 & 0.2979  \\
NGC3521  & 10.93 & 0.52 & 0.47 & 0.58 & 12.29 & 11.83 & 12.85 & 0.86 & 0.68 & 1.03 & 0.29  & 0.1212 & 0.0315 & 0.3787  \\
NGC3726  & 10.85 & 0.39 & 0.28 & 0.47 & 11.76 & 11.59 & 11.98 & 0.87 & 0.73 & 1.02 & 2.96  & 0.1987 & 0.1058 & 0.3469  \\
NGC3741  & 7.45 & 0.46 & 0.14 & 0.89 & 10.57 & 10.33 & 10.86 & 0.84 & 0.72 & 0.95 & 1.05  & 0.0013 & 0.0004 & 0.0031  \\
NGC3769  & 10.27 & 0.35 & 0.21 & 0.51 & 11.40 & 11.25 & 11.57 & 1.01 & 0.88 & 1.14 & 0.68  & 0.0970 & 0.0495 & 0.1719  \\
NGC3893  & 10.77 & 0.50 & 0.41 & 0.58 & 12.01 & 11.75 & 12.36 & 0.95 & 0.78 & 1.11 & 1.27  & 0.1227 & 0.0551 & 0.2310  \\
NGC3972  & 10.16 & 0.40 & 0.14 & 0.73 & 12.03 & 11.57 & 12.52 & 0.86 & 0.70 & 0.98 & 1.19  & 0.0240 & 0.0038 & 0.1250  \\
NGC3992  & 11.36 & 0.82 & 0.69 & 0.93 & 12.15 & 12.03 & 12.30 & 0.90 & 0.74 & 1.05 & 0.85  & 0.4339 & 0.3037 & 0.6160  \\
NGC4010  & 10.24 & 0.25 & 0.09 & 0.45 & 11.96 & 11.62 & 12.36 & 0.81 & 0.68 & 0.95 & 2.44  & 0.0216 & 0.0045 & 0.0764  \\
NGC4013  & 10.90 & 0.48 & 0.41 & 0.54 & 11.98 & 11.81 & 12.19 & 0.85 & 0.70 & 0.99 & 1.31  & 0.0776 & 0.0483 & 0.1161  \\
NGC4088  & 11.03 & 0.31 & 0.24 & 0.37 & 11.77 & 11.54 & 12.05 & 0.91 & 0.74 & 1.06 & 0.57  & 0.2924 & 0.1393 & 0.5817  \\
NGC4100  & 10.77 & 0.74 & 0.61 & 0.85 & 11.69 & 11.48 & 11.93 & 0.97 & 0.81 & 1.12 & 1.27  & 0.4199 & 0.2325 & 0.7423  \\
NGC4138  & 10.64 & 0.68 & 0.58 & 0.80 & 11.46 & 11.09 & 11.82 & 0.99 & 0.82 & 1.16 & 1.68  & 0.2491 & 0.1029 & 0.6693  \\
NGC4157  & 11.02 & 0.40 & 0.32 & 0.48 & 11.95 & 11.74 & 12.22 & 0.89 & 0.73 & 1.04 & 0.55  & 0.2388 & 0.1231 & 0.4311  \\
NGC4183  & 10.03 & 0.75 & 0.38 & 1.04 & 11.16 & 10.97 & 11.35 & 1.01 & 0.87 & 1.13 & 0.18  & 0.3102 & 0.1236 & 0.6488  \\
NGC4559  & 10.29 & 0.38 & 0.20 & 0.55 & 11.41 & 11.23 & 11.61 & 0.95 & 0.81 & 1.09 & 0.24  & 0.1513 & 0.0527 & 0.3176  \\
NGC5033  & 11.04 & 0.40 & 0.31 & 0.48 & 11.91 & 11.86 & 11.96 & 1.23 & 1.14 & 1.31 & 3.81  & 0.3049 & 0.1732 & 0.4383  \\
NGC5055  & 11.18 & 0.32 & 0.29 & 0.34 & 11.82 & 11.79 & 11.85 & 1.12 & 1.06 & 1.18 & 2.75  & 0.4220 & 0.3913 & 0.4514  \\
NGC5371  & 11.53 & 0.44 & 0.34 & 0.53 & 11.64 & 11.53 & 11.74 & 1.21 & 1.02 & 1.34 & 6.59  & 1.9570 & 1.1181 & 3.1110  \\
NGC5585  & 9.47 & 0.18 & 0.08 & 0.30 & 11.33 & 11.18 & 11.52 & 0.90 & 0.79 & 0.98 & 5.85  & 0.0142 & 0.0006 & 0.0294  \\
NGC5907  & 11.24 & 0.68 & 0.56 & 0.78 & 12.02 & 11.93 & 12.16 & 0.89 & 0.71 & 1.07 & 6.38  & 0.5110 & 0.4049 & 0.6183  \\
NGC5985  & 11.32 & 0.45 & 0.26 & 0.65 & 12.21 & 12.12 & 12.28 & 1.37 & 1.30 & 1.44 & 2.85  & 0.3156 & 0.1265 & 0.5595  \\
NGC6015  & 10.51 & 0.78 & 0.65 & 0.87 & 11.67 & 11.52 & 11.88 & 0.94 & 0.77 & 1.10 & 8.45  & 0.3054 & 0.1972 & 0.4377  \\
NGC6195  & 11.59 & 0.46 & 0.42 & 0.48 & 12.16 & 11.94 & 12.42 & 0.79 & 0.64 & 0.93 & 3.44  & 0.6961 & 0.3866 & 1.1779  \\
NGC6503  & 10.11 & 0.45 & 0.36 & 0.53 & 11.28 & 11.21 & 11.36 & 1.11 & 1.02 & 1.19 & 1.61  & 0.1585 & 0.1316 & 0.1883  \\
NGC6674  & 11.33 & 0.94 & 0.83 & 1.03 & 12.42 & 12.32 & 12.56 & 0.65 & 0.52 & 0.77 & 3.87  & 0.3996 & 0.2914 & 0.5274  \\
NGC6946  & 10.82 & 0.44 & 0.38 & 0.48 & 11.83 & 11.62 & 12.12 & 0.95 & 0.79 & 1.09 & 1.88  & 0.2336 & 0.1103 & 0.4250  \\
NGC7331  & 11.40 & 0.36 & 0.33 & 0.40 & 12.38 & 12.21 & 12.60 & 0.85 & 0.71 & 0.98 & 0.80  & 0.1527 & 0.0945 & 0.2232  \\
NGC7814  & 10.87 & 0.50 & 0.43 & 0.56 & 12.21 & 12.01 & 12.50 & 1.01 & 0.86 & 1.15 & 1.30  & 0.1245 & 0.0688 & 0.1869  \\
UGC00128  & 10.08 & 0.53 & 0.18 & 0.92 & 11.56 & 11.53 & 11.59 & 0.93 & 0.86 & 0.99 & 3.19  & 0.1058 & 0.0370 & 0.1797  \\
UGC00191  & 9.30 & 0.83 & 0.51 & 1.08 & 10.96 & 10.87 & 11.10 & 0.93 & 0.82 & 1.02 & 3.68  & 0.0947 & 0.0586 & 0.1368  \\
UGC00731  & 8.51 & 0.59 & 0.19 & 1.01 & 10.77 & 10.64 & 10.91 & 0.99 & 0.91 & 1.08 & 0.36  & 0.0176 & 0.0051 & 0.0338  \\
UGC02259  & 9.24 & 0.86 & 0.46 & 1.11 & 10.78 & 10.69 & 10.89 & 1.23 & 1.15 & 1.31 & 1.37  & 0.1220 & 0.0610 & 0.1851  \\
UGC02487  & 11.69 & 0.98 & 0.85 & 1.08 & 12.58 & 12.52 & 12.67 & 0.94 & 0.81 & 1.06 & 5.28  & 0.3968 & 0.3302 & 0.4704  \\
UGC02885  & 11.61 & 0.63 & 0.55 & 0.72 & 12.62 & 12.48 & 12.79 & 0.75 & 0.62 & 0.88 & 1.47  & 0.3448 & 0.2284 & 0.5073  \\
UGC02916  & 11.09 & 0.34 & 0.31 & 0.36 & 12.10 & 11.93 & 12.31 & 1.05 & 0.95 & 1.15 & 10.88  & 0.2354 & 0.1404 & 0.3645  \\
UGC02953  & 11.41 & 0.56 & 0.51 & 0.60 & 12.29 & 12.22 & 12.36 & 1.11 & 1.02 & 1.20 & 6.78  & 0.4796 & 0.3421 & 0.6312  \\
UGC03205  & 11.06 & 0.72 & 0.64 & 0.79 & 12.12 & 11.95 & 12.33 & 0.85 & 0.70 & 1.01 & 3.51  & 0.4040 & 0.2531 & 0.5862  \\
UGC03546  & 11.01 & 0.41 & 0.34 & 0.46 & 11.92 & 11.80 & 12.06 & 1.07 & 0.96 & 1.18 & 1.52  & 0.2236 & 0.1352 & 0.3344  \\
UGC03580  & 10.12 & 0.18 & 0.13 & 0.22 & 11.52 & 11.42 & 11.64 & 0.95 & 0.87 & 1.04 & 3.52  & 0.0459 & 0.0121 & 0.0823  \\
UGC04278  & 9.12 & 0.36 & 0.10 & 0.76 & 11.41 & 11.00 & 11.89 & 0.80 & 0.65 & 0.94 & 2.19  & 0.0095 & 0.0011 & 0.0430  \\
UGC04483  & 7.11 & 0.52 & 0.17 & 0.93 & 9.30 & 8.97 & 9.74 & 1.11 & 0.95 & 1.26 & 0.74  & 0.0160 & 0.0038 & 0.0485  \\
UGC04499  & 9.19 & 0.34 & 0.11 & 0.69 & 10.89 & 10.70 & 11.12 & 0.93 & 0.81 & 1.04 & 0.95  & 0.0322 & 0.0070 & 0.0839  \\
UGC05005  & 9.61 & 0.36 & 0.10 & 0.78 & 11.10 & 10.84 & 11.36 & 0.85 & 0.71 & 0.97 & 1.11  & 0.0718 & 0.0151 & 0.2207  \\
UGC05253  & 11.23 & 0.46 & 0.43 & 0.48 & 12.16 & 12.08 & 12.27 & 1.05 & 0.98 & 1.12 & 3.22  & 0.3759 & 0.2567 & 0.5165  \\
UGC05414  & 9.05 & 0.20 & 0.06 & 0.46 & 11.17 & 10.82 & 11.57 & 0.77 & 0.64 & 0.89 & 1.68  & 0.0061 & 0.0002 & 0.0256  \\
UGC05716  & 8.77 & 0.44 & 0.15 & 0.83 & 10.81 & 10.75 & 10.89 & 0.98 & 0.91 & 1.03 & 1.76  & 0.0186 & 0.0062 & 0.0312  \\
UGC05721  & 8.73 & 0.93 & 0.60 & 1.12 & 10.91 & 10.68 & 11.23 & 1.17 & 1.01 & 1.30 & 1.90  & 0.0317 & 0.0142 & 0.0596  \\
UGC05829  & 8.75 & 0.59 & 0.18 & 1.01 & 10.47 & 10.16 & 10.83 & 0.95 & 0.80 & 1.09 & 0.84  & 0.0539 & 0.0106 & 0.1593  \\
UGC05918  & 8.37 & 0.63 & 0.21 & 1.02 & 10.07 & 9.81 & 10.43 & 1.04 & 0.89 & 1.17 & 0.35  & 0.0580 & 0.0124 & 0.1611  \\
UGC06399  & 9.36 & 0.61 & 0.22 & 0.99 & 11.27 & 10.95 & 11.67 & 0.89 & 0.75 & 1.02 & 0.97  & 0.0362 & 0.0077 & 0.1135  \\
UGC06446  & 8.99 & 0.75 & 0.32 & 1.08 & 10.96 & 10.75 & 11.23 & 1.06 & 0.92 & 1.18 & 0.22  & 0.0385 & 0.0133 & 0.0808  \\
UGC06614  & 11.09 & 0.27 & 0.17 & 0.36 & 12.20 & 12.03 & 12.41 & 0.83 & 0.68 & 0.96 & 0.44  & 0.0828 & 0.0428 & 0.1474  \\
UGC06667  & 9.15 & 0.63 & 0.21 & 1.03 & 11.41 & 11.18 & 11.72 & 0.88 & 0.76 & 0.98 & 1.57  & 0.0113 & 0.0029 & 0.0275  \\
UGC06786  & 10.87 & 0.57 & 0.49 & 0.65 & 12.22 & 12.10 & 12.37 & 1.05 & 0.94 & 1.16 & 1.47  & 0.1669 & 0.1166 & 0.2240  \\
UGC06787  & 10.99 & 0.43 & 0.38 & 0.47 & 12.17 & 12.10 & 12.24 & 1.19 & 1.12 & 1.26 & 27.20  & 0.2041 & 0.1410 & 0.2737  \\
UGC06917  & 9.83 & 0.46 & 0.18 & 0.78 & 11.46 & 11.23 & 11.77 & 0.93 & 0.79 & 1.05 & 0.75  & 0.0438 & 0.0137 & 0.1163  \\
UGC06923  & 9.46 & 0.30 & 0.11 & 0.59 & 11.20 & 10.83 & 11.68 & 0.94 & 0.78 & 1.08 & 0.85  & 0.0194 & 0.0035 & 0.0809  \\
UGC06930  & 9.95 & 0.68 & 0.28 & 1.02 & 11.15 & 10.93 & 11.38 & 0.99 & 0.86 & 1.12 & 0.33  & 0.2057 & 0.0617 & 0.4919  \\
UGC06973  & 10.73 & 0.18 & 0.16 & 0.20 & 12.83 & 12.24 & 13.53 & 0.86 & 0.65 & 1.06 & 1.11  & 0.0032 & 0.0006 & 0.0126  \\
UGC06983  & 9.72 & 0.76 & 0.38 & 1.06 & 11.31 & 11.11 & 11.57 & 1.00 & 0.85 & 1.13 & 0.70  & 0.0767 & 0.0301 & 0.1557  \\
UGC07089  & 9.55 & 0.44 & 0.13 & 1.05 & 10.68 & 9.71 & 11.15 & 0.91 & 0.75 & 1.13 & 1.01  & 0.1587 & 0.0203 & 3.9876  \\
UGC07125  & 9.43 & 0.28 & 0.09 & 0.57 & 10.46 & 10.33 & 10.60 & 0.91 & 0.81 & 1.01 & 1.08  & 0.1392 & 0.0239 & 0.3678  \\
UGC07151  & 9.36 & 0.84 & 0.58 & 1.06 & 10.77 & 10.45 & 11.14 & 0.95 & 0.80 & 1.07 & 2.64  & 0.1613 & 0.0586 & 0.4149  \\
UGC07399  & 9.06 & 0.84 & 0.45 & 1.10 & 11.39 & 11.17 & 11.70 & 1.13 & 1.01 & 1.23 & 1.74  & 0.0163 & 0.0066 & 0.0315  \\
UGC07524  & 9.39 & 0.50 & 0.17 & 0.94 & 11.00 & 10.77 & 11.27 & 0.87 & 0.75 & 0.97 & 0.94  & 0.0657 & 0.0155 & 0.1930  \\
UGC07559  & 8.04 & 0.53 & 0.15 & 1.02 & 9.31 & 8.70 & 9.76 & 1.08 & 0.92 & 1.27 & 1.29  & 0.1263 & 0.0185 & 1.0510  \\
UGC07603  & 8.58 & 0.53 & 0.20 & 0.88 & 11.01 & 10.70 & 11.44 & 0.97 & 0.82 & 1.11 & 1.62  & 0.0084 & 0.0021 & 0.0224  \\
UGC07690  & 8.93 & 0.89 & 0.66 & 1.08 & 10.18 & 9.87 & 10.53 & 1.09 & 0.94 & 1.25 & 0.48  & 0.1986 & 0.0754 & 0.4751  \\
UGC07866  & 8.09 & 0.66 & 0.22 & 1.06 & 9.31 & 8.78 & 9.80 & 1.14 & 0.97 & 1.30 & 0.23  & 0.1754 & 0.0266 & 0.9472  \\
UGC08286  & 9.10 & 0.94 & 0.61 & 1.13 & 10.90 & 10.78 & 11.05 & 1.11 & 1.02 & 1.20 & 2.13  & 0.0801 & 0.0490 & 0.1160  \\
UGC08490  & 9.01 & 0.92 & 0.58 & 1.12 & 10.79 & 10.64 & 10.99 & 1.15 & 1.01 & 1.27 & 0.29  & 0.0746 & 0.0425 & 0.1147  \\
UGC08550  & 8.46 & 0.79 & 0.42 & 1.07 & 10.51 & 10.33 & 10.74 & 1.05 & 0.93 & 1.16 & 0.66  & 0.0314 & 0.0154 & 0.0546  \\
UGC08699  & 10.70 & 0.56 & 0.51 & 0.60 & 11.95 & 11.75 & 12.21 & 0.99 & 0.85 & 1.11 & 1.13  & 0.1982 & 0.1076 & 0.3284  \\
UGC09037  & 10.84 & 0.11 & 0.04 & 0.20 & 11.91 & 11.74 & 12.13 & 0.87 & 0.74 & 0.98 & 1.03  & 0.0381 & 0.0101 & 0.0852  \\
UGC09133  & 11.45 & 0.47 & 0.44 & 0.50 & 12.22 & 12.18 & 12.25 & 0.99 & 0.92 & 1.05 & 8.84  & 0.5423 & 0.4231 & 0.6673  \\
UGC10310  & 9.24 & 0.73 & 0.30 & 1.06 & 10.67 & 10.42 & 10.96 & 1.02 & 0.88 & 1.14 & 0.49  & 0.1258 & 0.0341 & 0.3281  \\
UGC11820  & 8.99 & 0.52 & 0.17 & 0.90 & 11.15 & 11.04 & 11.28 & 0.74 & 0.65 & 0.81 & 2.20  & 0.0221 & 0.0079 & 0.0377  \\
UGC11914  & 11.18 & 0.64 & 0.61 & 0.67 & 13.04 & 12.44 & 13.67 & 0.75 & 0.58 & 0.94 & 2.55  & 0.0492 & 0.0110 & 0.2009  \\
UGC12506  & 11.14 & 0.97 & 0.66 & 1.14 & 12.14 & 11.96 & 12.33 & 0.99 & 0.84 & 1.13 & 0.67  & 0.5698 & 0.2753 & 0.9742  \\
UGC12632  & 9.11 & 0.66 & 0.23 & 1.04 & 10.73 & 10.56 & 10.92 & 0.98 & 0.87 & 1.09 & 0.41  & 0.0878 & 0.0252 & 0.1817  \\
UGC12732  & 9.22 & 0.54 & 0.18 & 0.95 & 11.11 & 10.96 & 11.30 & 0.92 & 0.80 & 1.02 & 0.29  & 0.0361 & 0.0109 & 0.0741  \\
UGCA281  & 8.29 & 0.66 & 0.28 & 1.01 & 9.86 & 9.36 & 10.46 & 1.04 & 0.88 & 1.18 & 0.89  & 0.0382 & 0.0054 & 0.1712  \\
UGCA444  & 7.08 & 0.61 & 0.21 & 1.02 & 9.62 & 9.19 & 10.14 & 1.08 & 0.91 & 1.25 & 0.55  & 0.0088 & 0.0018 & 0.0316  \\
\end{longtable}
\end{center}
\end{landscape}
\clearpage
\twocolumn

%------ FIG. 3 --------------------------------------------------------------------------------
\begin{figure*}
\includegraphics[width=\textwidth]{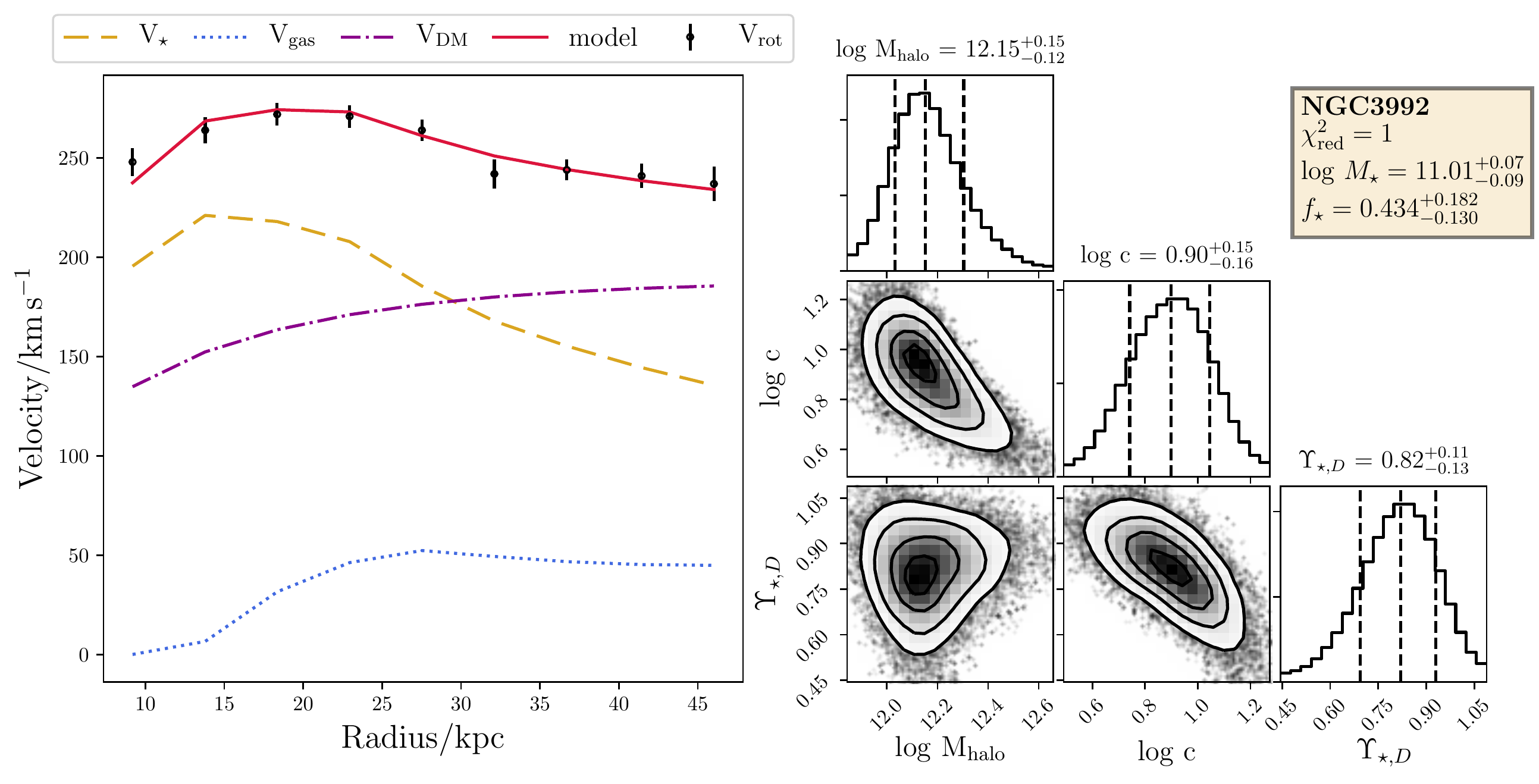}
\caption{Example of rotation curve decomposition for NGC 3992. In the \emph{left-hand panel}, we
         show the observed rotation curve (black points) with our best model (red solid curve),
         which we also decompose into the contributions from stars (gold dashed curve), gas
         (blue dotted curve) and dark matter (purple dot-dashed curve). In the \emph{right-hand panel},
         we show the posterior distributions of the three parameters of the model: halo mass, halo
         concentration and mass-to-light ratio of the stellar disc. Similar plots for all the other
         galaxies in our sample can be found online at
         \url{http://astro.u-strasbg.fr/~posti/PFM19_fiducial_fits/}.}
\label{fig:curve_decomp}
\end{figure*}

\begin{figure*}
\includegraphics[width=0.33\textwidth]{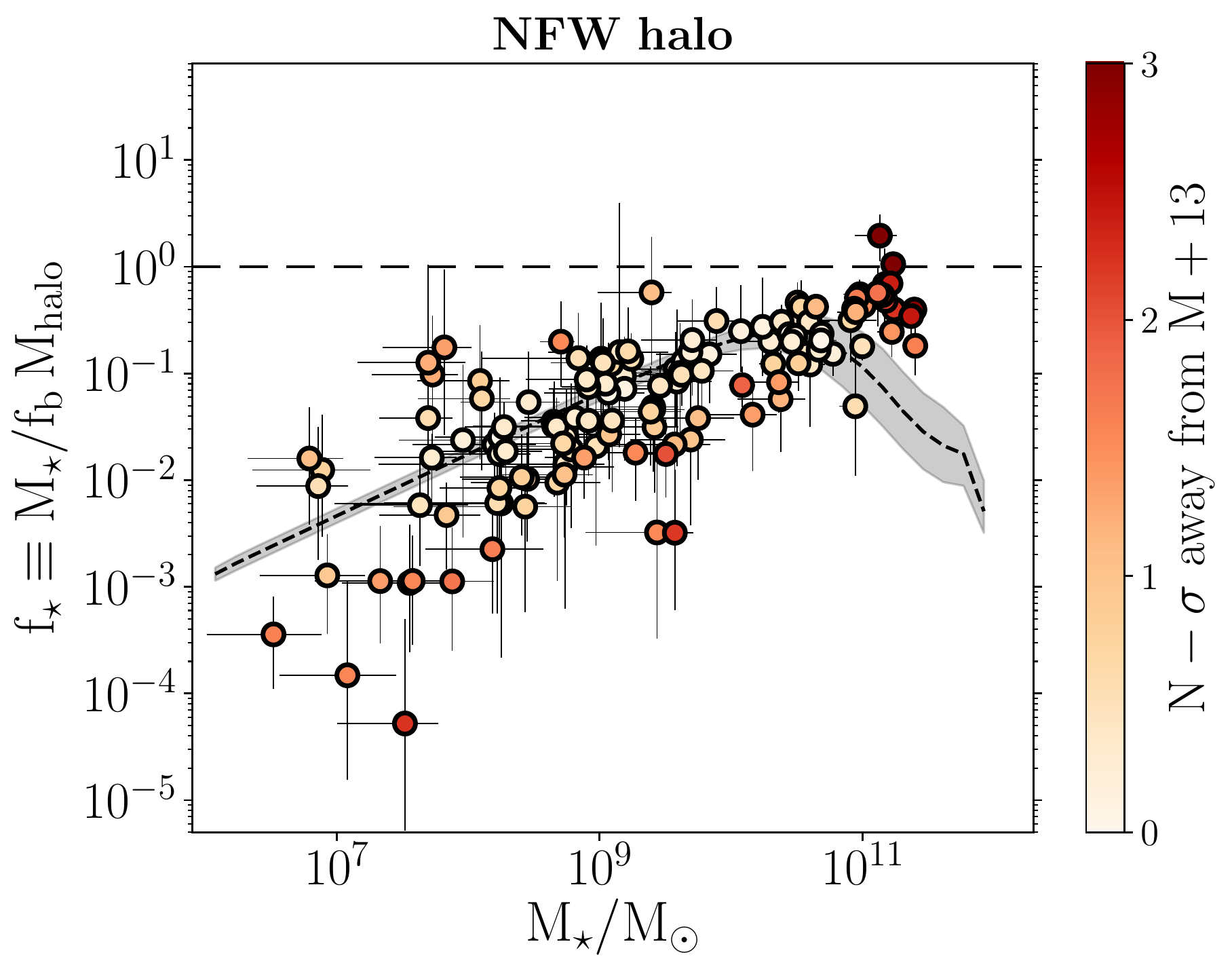}
\includegraphics[width=0.33\textwidth]{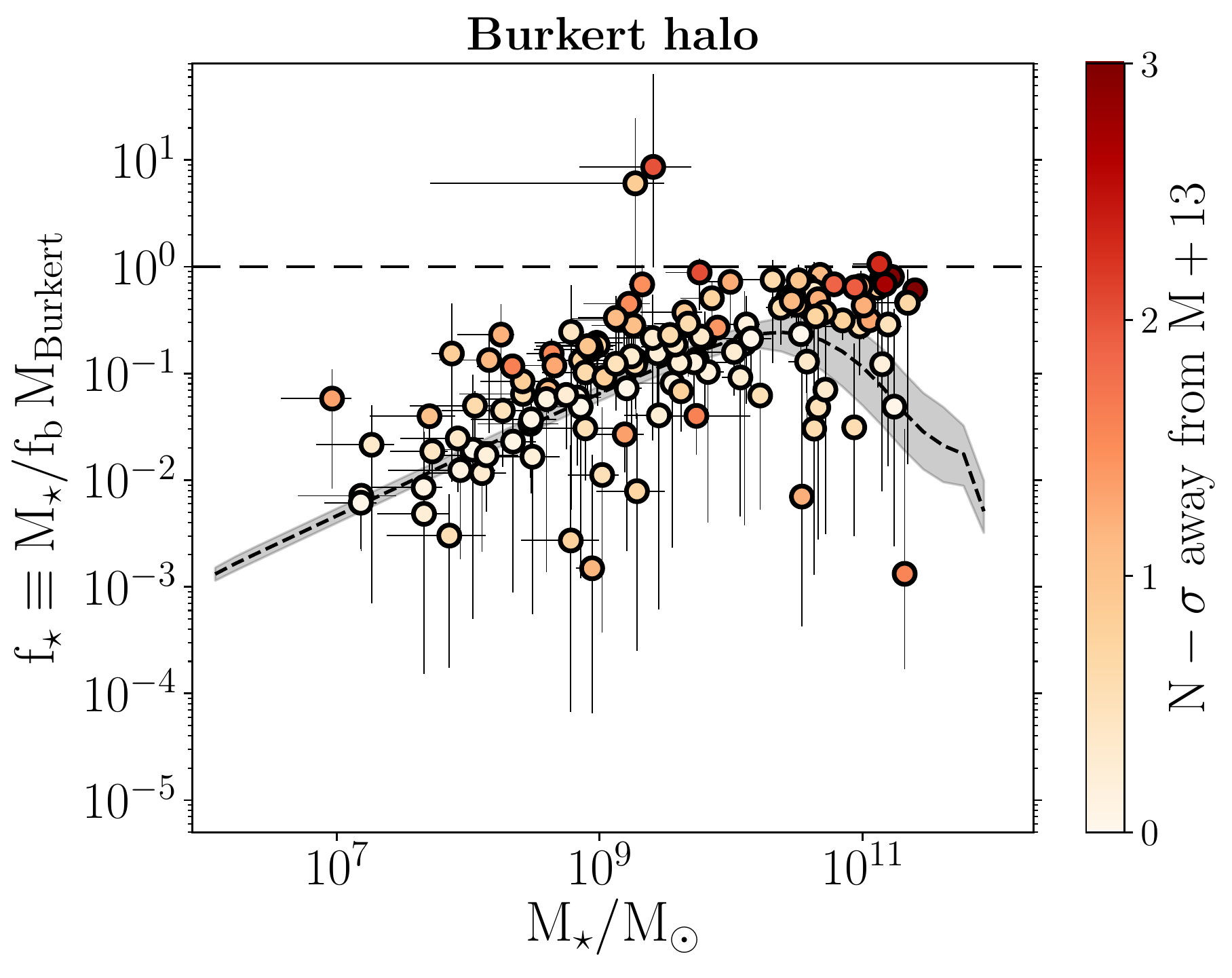}
\includegraphics[width=0.33\textwidth]{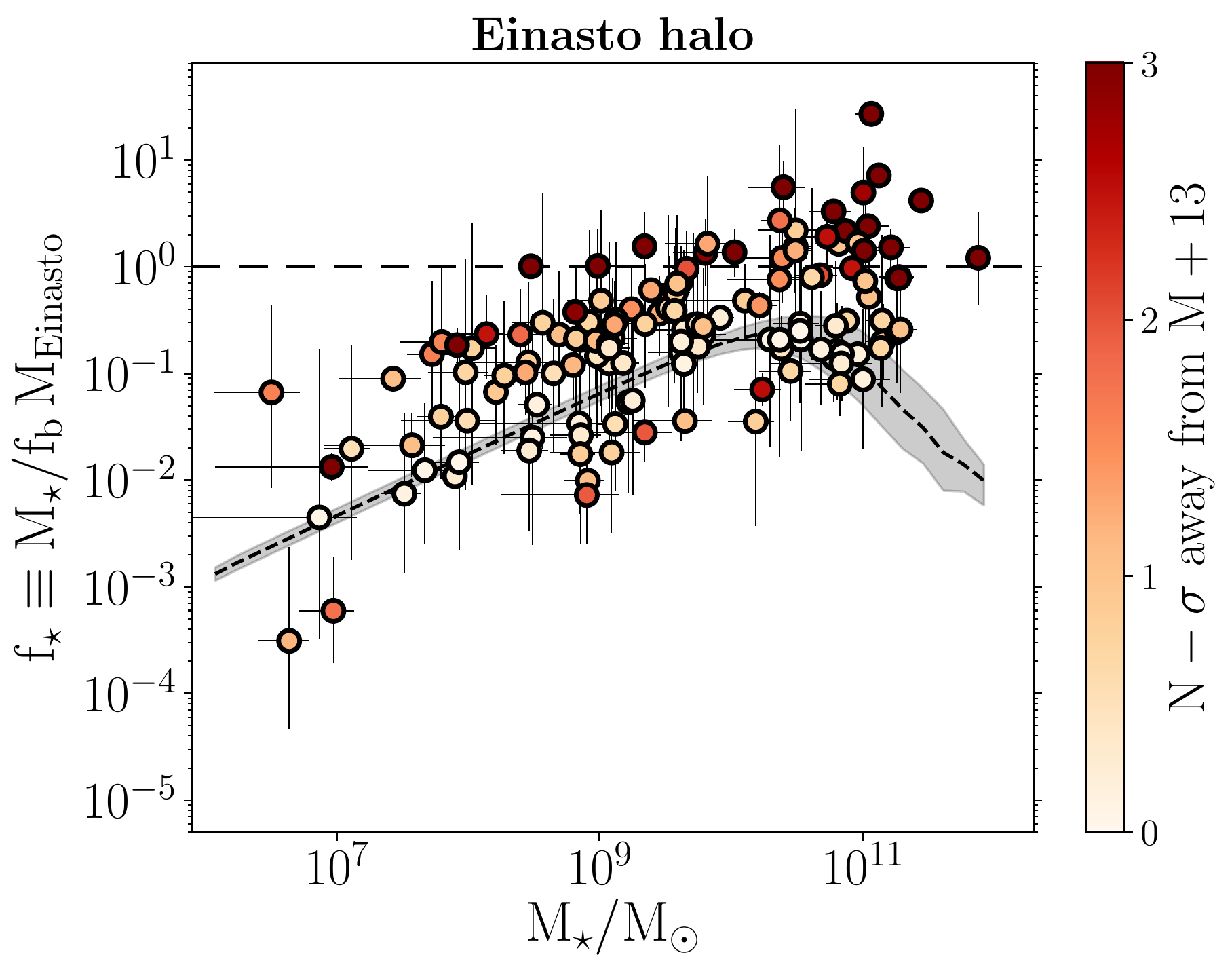} \\
\includegraphics[width=0.33\textwidth]{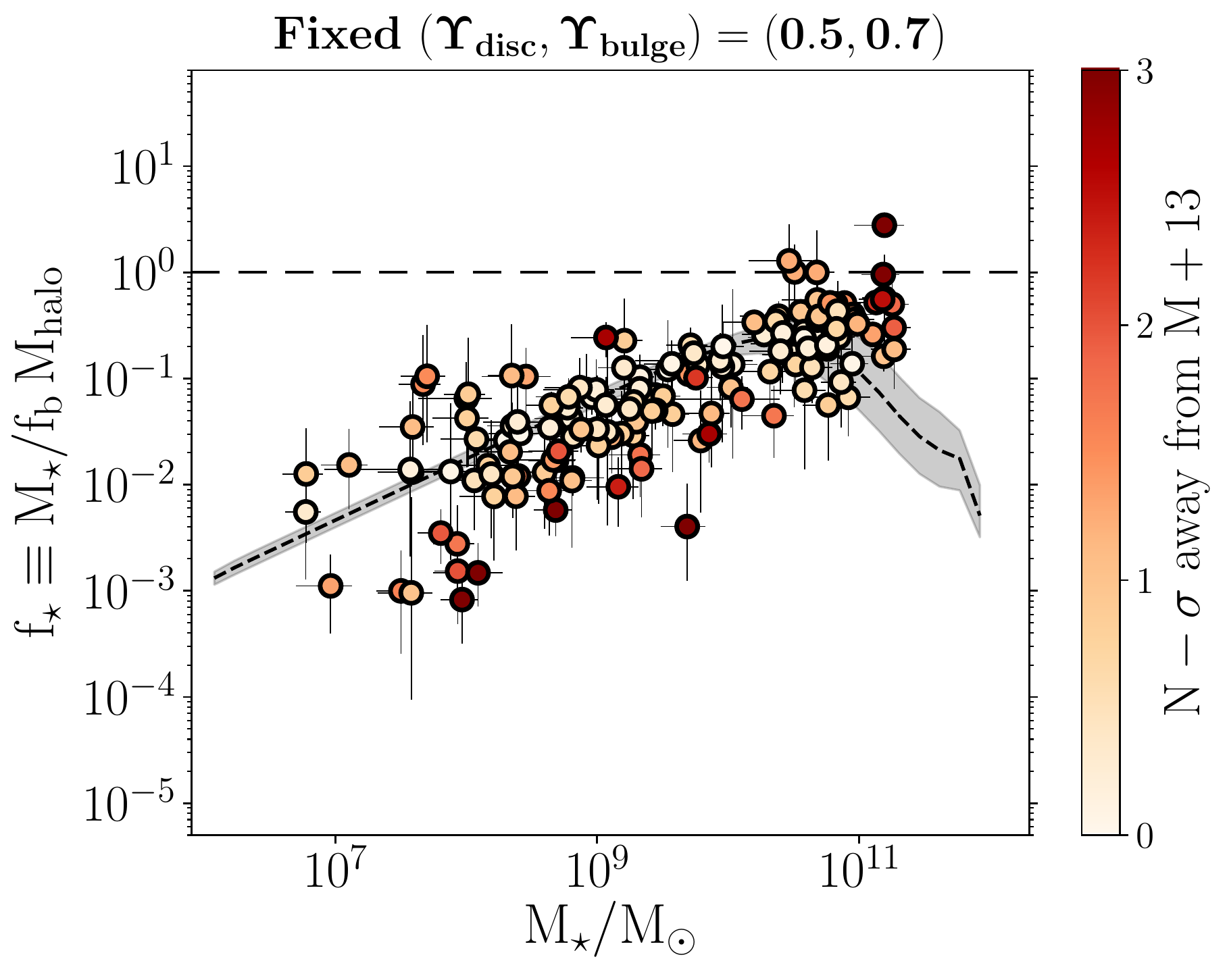}
\includegraphics[width=0.33\textwidth]{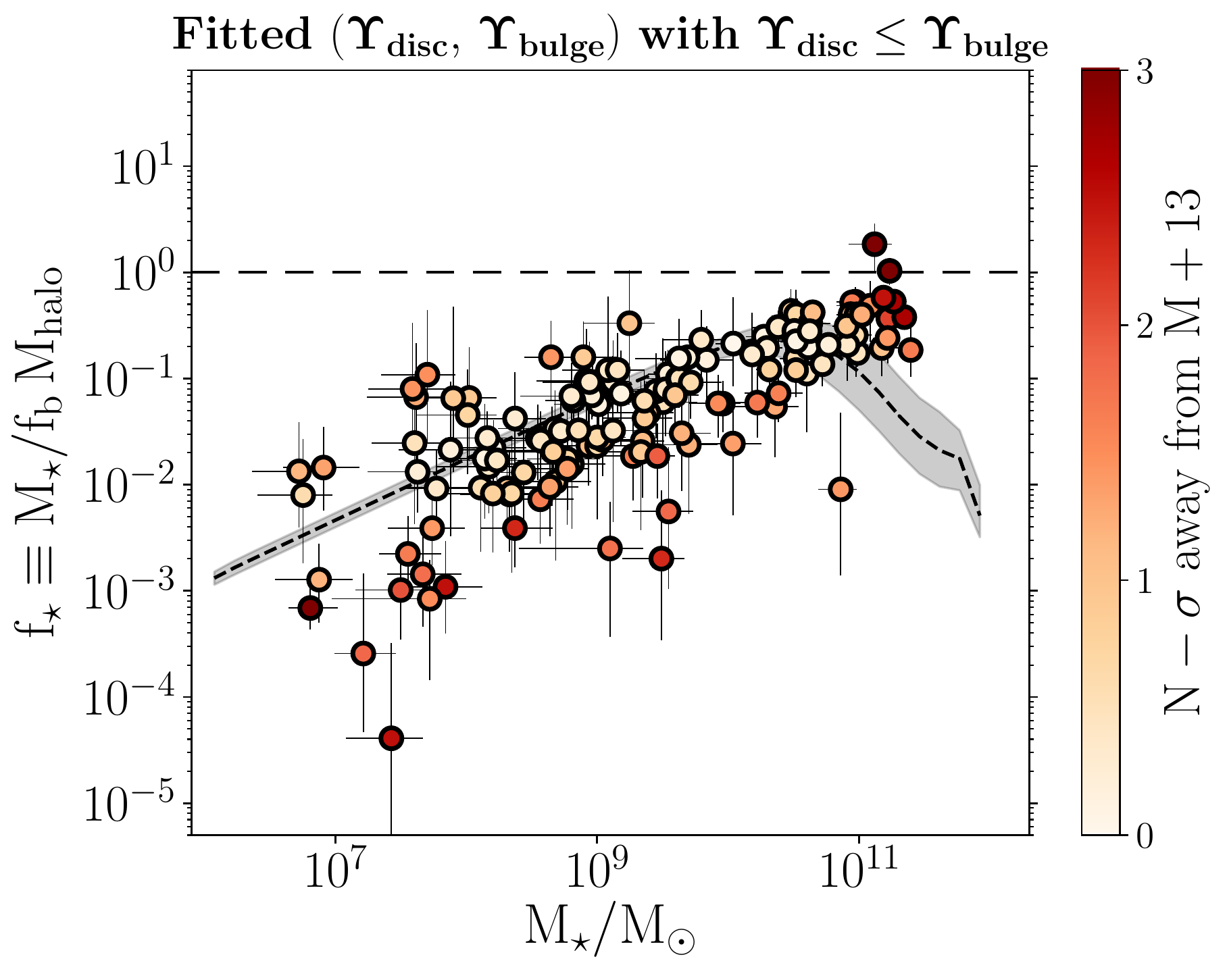}
\caption{Resulting $\fstar-\Mstar$ relation when varying the assumptions on the fit of the galaxy
         rotation curves. In the top row we varied the dark matter halo model: NFW (left), Burkert
         (centre) or Einasto (right). In the first two cases, we have fitted the rotation curves
         with a uniform prior on $\Udisc$, assuming $\Ubulge=1.4\Udisc$ and with a prior on the
         concentration-mass relation for the NFW profile \citep[from][]{DuttonMaccio14} and one on
         the core radius-core mass relation for the Burkert profile \citep[from][]{SalucciBurkert00}.
         The fits in the Einasto case are instead obtained by \cite{Ghari+19}, who used the
         mass-to-light ratios derived by \cite{Li+18}.
         In the bottom row, we show the cases where we used an NFW halo, but varied the assumptions
         on the mass-to-light ratios: either we fixed them (left) or we left both of them free to
         vary with the condition $\Udisc\leq\Ubulge$ (right). In all panels the colouring of the
         points, the dashed horizontal line and the abundance matching predictions (dashed curve with
         grey band) are as in Fig.~\ref{fig:mstar_fstar}.}
\label{fig:tests}
\end{figure*}

\end{document}